\documentclass[11pt,letterpaper]{JHEP3}
\usepackage{amsmath,epsfig,array}
\usepackage{graphicx}
\usepackage{amssymb}
\usepackage{amsfonts,amstext,latexsym,xspace}
\usepackage{amsfonts}
\usepackage{amsmath}
\usepackage{cite}

\makeatletter
\newif\if@fewtab\@fewtabtrue
\def\moth{\mathsurround=0pt}
\newdimen\zo \zo=0pt

\def\tick{\leaders\hrule height 0.5ex depth 0pt \hskip 0.5pt}
\def\upboxfill{$\moth \setbox\zo\hbox{\tick}%
  \hskip 2pt\hbox to 0pt{$\tick$\hss}\hrulefill \hbox to 2pt{$\tick$\hss}$}

\def\dtick{\leaders\hrule height .34pt depth 0.5ex \hskip 0.5pt}
\def\downboxfill{$\moth \setbox\zo\hbox{\dtick}%
  \hskip 2pt\hbox to 0pt{$\dtick$\hss}\hrulefill%
  \hbox to 2pt{$\dtick$\hss}$}

\usepackage{array}
\usepackage{amsmath}
\usepackage{amssymb}
\usepackage{graphicx}
\usepackage{longtable}

\newcommand{\commute}[2]{\left[ #1 \, , \, #2 \right]}
\newcommand{\anticommute}[2]{\left\{ #1 \, , \, #2 \right\}}
\newcommand{\ket}[1]{\left\lvert #1 \right\rangle}
\newcommand{\eq}{\begin{equation}}
\newcommand{\en}{\end{equation}}
\newcommand{\eqq}{\begin{eqnarray}}
\newcommand{\enn}{\end{eqnarray}}

\newcommand{\stln}{\setlength{\unitlength}{10pt}}  
\newcommand{\fr}{\framebox(1,1){}}  
\newcommand{\lfr}{\framebox(2,1){}}  
\newcommand{\sfr}{\framebox(1,1)[bl]{\begin{picture}(1,1)(0,0)
                                      \put(0,0){\line(1,1){1}}
                                     \end{picture}
                                     }
                  }  
\newcommand{\dotfr}{\stln \lower2.0pt \hbox{\begin{picture}(2,1)(0,0)
                                                 \put(0,0){\lfr}
                                                 \put(0.5,0.5){\dots}
                                                \end{picture}
                                                }
                        } 
\newcommand{\sonebox}{\stln \lower2.0pt \hbox{\begin{picture}(1,1)(0,0)
                                               \put(0,0){\sfr}
                                              \end{picture}
                                              }
                      }
\newcommand{\stwobox}{\stln \lower2.0pt \hbox{\begin{picture}(2,1)(0,0)
                                               \multiput(0,0)(1,0){2}{\sfr}
                                              \end{picture}
                                              }
                      }
\newcommand{\sgenrowbox}{\stln \lower2.0pt \hbox{\begin{picture}(5,1)(0,0)
                                               \multiput(0,0)(1,0){2}{\sfr}
                                               \put(2,0.2){\dotfr}
                                               \put(4,0){\sfr}
                                              \end{picture}
                                              }
                      }
\newcommand{\genrowbox}{\stln \lower2.0pt \hbox{\begin{picture}(5,1)(0,0)
                                               \multiput(0,0)(1,0){2}{\fr}
                                               \put(2,0.2){\dotfr}
                                               \put(4,0){\fr}
                                              \end{picture}
                                              }
                      }

\title{$SU(2)$ deformations of the minimal unitary representation of $OSp(8^*|2N)$ as massless $6D$ conformal supermultiplets}
\author{
Sudarshan Fernando$^{1}$\footnote{fernando@kutztown.edu}  and
Murat G\"{u}naydin$^{2}$\footnote{murat@phys.psu.edu}
\\

$^{1}$\emph{Physical Sciences Department\\Kutztown University\\ Kutztown, PA 19530, USA}  \\

$^{2}$\emph{Center for Fundamental Theory \\ Institute for Gravitation and the Cosmos \\ Physics Department \\
Pennsylvania State University\\
University Park, PA 16802, USA} }

\abstract{Minimal unitary representation of  $SO^*(8) \simeq SO(6,2)$ realized over the Hilbert space of functions of five variables and its deformations labeled by the spin $t$ of an $SU(2)$ subgroup correspond to massless conformal fields in six dimensions as was shown in {\tt arXiv:1005.3580}. In this paper we study the minimal unitary supermultiplet of  $OSp(8^*|2N)$ with the even subgroup $SO^*(8) \times USp(2N)$ and its  deformations using quasiconformal methods.
We show that the minimal unitary supermultiplet of $OSp(8^*|2N)$ admits deformations labeled uniquely  by the spin $t$ of an $SU(2)$ subgroup of  the little group $SO(4)$ of lightlike vectors in six dimensions. We construct the deformed minimal unitary representations and show that they correspond to massless $6D$ conformal supermultiplets.  The minimal unitary supermultiplet of $OSp(8^*|4)$
is the massless supermultiplet of $(2,0)$ conformal field theory that is
believed to be dual to M-theory on $AdS_7 \times S^4$. We study its deformations in further detail and show that they  are isomorphic to the doubleton supermultiplets constructed by using twistorial oscillators. }


\begin{document}


\section{Introduction}
\label{Intro}

Motivated by the problem of  constructing  the relevant unitary representations  of noncompact U-duality groups of extended supergravity theories, a general oscillator method was developed in  \cite{Gunaydin:1981dc,Gunaydin:1981yq,Gunaydin:1981zm}. The method, as formulated in  \cite{Gunaydin:1981yq},  generalized and  unified  the  special  constructions  that had previously appeared in the physics literature.   The general oscillator construction was later extended to  noncompact supergroups in \cite{Bars:1982ep} using bosonic as well as fermionic oscillators. In the generalized oscillator constructions  of \cite{Gunaydin:1981yq} and \cite{Bars:1982ep}, one realizes the generators of noncompact groups or supergroups  as bilinears of an arbitrary number  $P^C$ (``colors'') of  sets of oscillators transforming in a definite  representation (typically fundamental) of their maximal compact subgroups or subsupergroups.
  For symplectic groups $Sp(2n,\mathbb{R})$, the minimum  value of $P^C$ is  one, and the resulting unitary representations are  the singleton representations, which are referred to as metaplectic representations in the mathematics literature.  Symplectic groups  $Sp(2n,\mathbb{R})$ admit only two singleton irreducible representations (irreps).  In general, the minimum allowed value of $P^C$  is two, and the resulting unitary representations of such noncompact groups were later called doubleton representations. For example, for the groups $SU(n,m)$ and $SO^*(2n)$, with maximal compact subgroups $SU(m)\times SU(n) \times U(1)$ and $U(n)$, respectively, one finds that $P^C_{min}=2$.  When $P^C_{min}=2$,  the noncompact group or supergroup  admits   an infinite number of doubleton irreps.
  The positive energy singleton or doubleton irreps  of noncompact groups or supergroups  do not belong to the discrete series representations. However, by tensoring them, one obtains positive energy unitary representations that, in general, belong to the holomorphic discrete series representations  of the respective noncompact group or supergroup.

The Kaluza-Klein spectrum of IIB  supergravity spontaneously compactified over the product space  $AdS_5 \times S^5$  of $5D$ anti-de Sitter space $AdS_5$ with the five sphere $S^5$ was first  obtained via the oscillator method by simple tensoring of the CPT self-conjugate doubleton supermultiplet of  $N=8$ $AdS_5$ superalgebra $PSU(2,2\,|\,4)$ repeatedly with itself and restricting to the CPT self-conjugate short supermultiplets of $PSU(2,2\,|\,4)$ \cite{Gunaydin:1984fk}. The CPT self-conjugate doubleton supermultiplet of the symmetry  superalgebra $PSU(2,2\,|\,4)$ of $AdS_5 \times S^5$  solution of IIB supergravity does not have a Poincar\'e limit in five dimensions and decouples from the Kaluza-Klein spectrum as gauge modes. This led the authors of \cite{Gunaydin:1984fk} to  propose that the  field theory of CPT self-conjugate doubleton supermultiplet  of $PSU(2,2\,|\,4)$  must live on the boundary of $AdS_5$, which can be identified with $4D$ Minkowski space on which $SO(4,2)$ acts as a conformal group, and the unique candidate for this theory is the  four dimensional $N=4$ super Yang-Mills theory that was known to be conformally invariant.

The spectra of  spontaneous compactifications of eleven dimensional supergravity over $AdS_4 \times S^7$ and $AdS_7 \times S^4$, that had been obtained by other methods previously, were fitted into supermultiplets of the  symmetry superalgebras $OSp(8\,|\,4,\mathbb{R})$ and $OSp(8^*|4)$ constructed  by oscillator methods in \cite{Gunaydin:1985tc} and \cite{Gunaydin:1984wc}, respectively. Furthermore, the entire Kaluza-Klein spectra of eleven dimensional supergravity over these two spaces were obtained by tensoring the singleton and scalar doubleton supermultiplets of $OSp(8\,|\,4,\mathbb{R})$ and $OSp(8^*|4)$, respectively.
The relevant singleton supermultiplet of $OSp(8\,|\,4,\mathbb{R})$ and scalar doubleton supermultiplet of $OSp(8^*|4)$  do not have a Poincar\'e limit in four and seven dimensions, respectively,   and decouple from the respective spectra as gauge modes. Again it was proposed that the field theories of the singleton and scalar doubleton supermutiplets live on the boundaries of $AdS_4$ and $AdS_7$ as superconformally invariant theories \cite{Gunaydin:1984wc,Gunaydin:1985tc}.

 These results have become an integral  part of the work on AdS/CFT dualities in M-/superstring theory since the famous paper of  Maldacena\cite{Maldacena:1997re} and subsequent works  of  Witten \cite{Witten:1998qj}  and of Gubser et al. \cite{Gubser:1998bc}.

 Noncompact groups entered   physics  as spectrum generating symmetry groups during the 1960s. The work of  physicists on spectrum generating symmetry groups motivated  Joseph to introduce the  concept of  minimal unitary representations of Lie groups  in \cite{MR0342049}.
These are   defined as unitary
representations of  corresponding noncompact groups over  Hilbert spaces
of functions of  smallest possible (minimal) number of variables.  Joseph gave the
minimal realizations of the complex forms of classical Lie algebras
and of the exceptional Lie algebra $\mathfrak{g}_2$ in a Cartan-Weyl basis.   The
  minimal unitary representation of the split exceptional group $E_{8(8)}$ was first identified within  Langland's classification  by Vogan
\cite{MR644845}. Later, Kostant studied  the minimal unitary representation of $SO(4,4)$ and its relation to triality in    \cite{MR1103588}.
A general study of  minimal unitary representations of
simply laced groups   was given  by Kazhdan and Savin\cite{MR1159103}
and by Brylinski and Kostant
\cite{MR1372999,MR1278630}. Pioline, Kazhdan and Waldron \cite{Kazhdan:2001nx}
reformulated  the minimal unitary representations of simply laced
groups given in \cite{MR1159103} and  determined
the spherical vectors for the simply laced exceptional groups.
The minimal unitary representations of quaternionic real forms of exceptional Lie
groups were  studied by Gross and Wallach in \cite{MR1327538} and those
of $SO(p,q)$  in \cite{MR1108044,MR2020550,MR2020551,MR2020552}.
 The relation of minimal representations of $SO(p,q)$ to conformal geometry was studied rather recently  in \cite{Gover:2009vc}.

Over the last decade, there has been a great deal of progress  made towards the goal of constructing physically relevant unitary representations  of U-duality groups of extended supergravity theories.  This was partly motivated by the proposals that certain extensions of U-duality groups  act as spectrum generating symmetry groups of extremal black hole solutions in these theories.
For example, the classification of the
orbits of extremal black hole solutions in $N=8$ supergravity and
$N=2$ Maxwell-Einstein supergravity theories with symmetric scalar
manifolds led to the proposal that four dimensional U-duality groups act
as spectrum generating conformal symmetry groups of  corresponding
five dimensional supergravity theories
\cite{Ferrara:1997uz,Gunaydin:2000xr,Gunaydin:2004ku,Gunaydin:2003qm,Gunaydin:2005gd,Gunaydin:2009pk}.   Extension of this proposal to  corresponding spectrum generating symmetry groups of  extremal black hole solutions of four dimensional supergravity theories with symmetric scalar manifolds led to the discovery of novel geometric quasiconformal realizations  of three dimensional U-duality groups  \cite{Gunaydin:2000xr}. Quasiconformal extensions of four dimensional U-duality groups were then proposed   as spectrum generating symmetry groups of the
corresponding supergravity theories
\cite{Gunaydin:2000xr,Gunaydin:2004ku,Gunaydin:2003qm,Gunaydin:2005gd,Gunaydin:2009pk}. A concrete  framework for the implementation of the proposal that three
dimensional U-duality groups act as spectrum generating quasiconformal
groups was given in
\cite{Gunaydin:2005mx,Gunaydin:2007bg,Gunaydin:2007qq}. This framework was based on  the equivalence of equations of attractor flows of spherically
symmetric stationary BPS black holes of four dimensional supergravity
theories and the geodesic equations of a fiducial particle moving in the target space of  three dimensional supergravity theories obtained by reduction of the $4D$ theories  on a timelike circle \cite{Breitenlohner:1987dg}.

Quasiconformal realization of three dimensional  U-duality group $E_{8(8)}$ of maximal supergravity
in three dimensions is the first known geometric realization of any real form of
$E_{8}$\cite{Gunaydin:2000xr}.  As a quasiconformal group the action of $E_{8(8)}$  leaves invariant
a generalized light-cone with respect to a quartic distance function  in 57
dimensions. Quasiconformal realizations exist for various  real forms of all noncompact groups as well as for their complex forms  \cite{Gunaydin:2000xr,Gunaydin:2005zz}.
Furthermore, the quantization of  geometric quasiconformal action of a noncompact group leads directly to its minimal unitary representation.  This  was first shown  explicitly for the  maximally split
exceptional group  $E_{8(8)}$ with the maximal compact subgroup $SO(16)$
\cite{Gunaydin:2001bt} and  for the   three dimensional U-duality group $E_{8(-24)}$  of
the exceptional supergravity theory  \cite{Gunaydin:1983rk}
   in \cite{Gunaydin:2004md}.
The minimal unitary representations of  U-duality groups $F_{4(4)}$, $E_{6(2)}$,
 $E_{7(-5)}$ ,  $E_{8(-24)}$ and $SO(d+2,4)$ of $3D$ $N=2$ Maxwell-Einstein supergravity theories with symmetric scalar manifolds were studied in
\cite{Gunaydin:2005zz,Gunaydin:2004md}.
   A unified
formulation of the minimal unitary representations of certain noncompact real forms of
 groups  of type $A_2$, $G_2$, $D_4$, $F_4$, $E_6$, $E_7$,
$E_8$ and $C_n$ was given
 in \cite{Gunaydin:2006vz}. The minimal unitary
representations of $Sp\left(2n,\mathbb{R}\right)$ are simply the
singleton representations. In \cite{Gunaydin:2006vz},  minimal unitary representations of
 noncompact groups $SU\left(m,n\right)$, $SO\left(m,n\right)$,
$SO^*(2n)$ and $SL\left(m,\mathbb{R}\right)$ obtained by quantization of their quasiconformal realizations  were also given
explicitly.
Furthermore, this unified  approach was generalized to define and construct the corresponding
minimal representations of non-compact supergroups $G$ whose even
subgroups are of the form $H\times SL(2,\mathbb{R})$ with $H$
compact.

In mathematics literature, the term minimal unitary representation, in general, refers to a unique  representation of a  noncompact group. Symplectic groups $Sp(2N,\mathbb{R})$ admit two singleton irreps whose quadratic Casimirs take on the same value. They are both minimal unitary representations,  though in some of the mathematics literature only the scalar singleton is referred to as the minrep.   Similarly  the supergroups $OSp(M\,|\,2N,\mathbb{R}) $ with the even subgroup $SO(M) \times Sp(2N,\mathbb{R})$ admit two inequivalent singleton supermultiplets \cite{Gunaydin:1985tc,Gunaydin:1988kz,Gunaydin:1987hb}.
For noncompact groups or supergroups that admit only doubleton irreps, this raises the question as to whether any of the doubleton unitary representations can be identified with the  minimal unitary representation, and if so, how the infinite set of doubletons are related to the minrep.
This question was addressed for $5D$ anti-de Sitter  or $4D$ conformal group $SU(2,2)$ and  corresponding supergroups $SU(2,2\,|\,N)$ in our earlier work \cite{Fernando:2009fq}. We
showed that  the minimal unitary representation of the group $SU(2,2)$ obtained  by quantization of its quasiconformal realization coincides with the scalar doubleton representation corresponding to a massless scalar field in four dimensions. Furthermore  the minrep of $SU(2,2)$ admits a one-parameter ($\zeta$) family of deformations, and for a positive
(negative)  integer value of the deformation parameter $\zeta$,  one obtains
a positive energy
unitary irreducible representation of  $SU(2,2)$ corresponding to a
massless conformal field in four dimensions  transforming in  $\left( 0
\,,\, \frac{\zeta}{2} \right)$ $\left( \left( -\frac{\zeta}{2} \,,\, 0
\right) \right)$ representation of the Lorentz subgroup, $SL(2,\mathbb{C})$. We showed that these representations  are simply the doubletons  of $SU(2,2)$  that describe massless conformal fields in four dimensions \cite{Gunaydin:1998sw,Gunaydin:1998jc}. They were referred to as ladder (or most degenerate discrete series) unitary representations in some of the earlier literature on conformal group and it was shown  by Mack and Todorov that they remain irreducible under restriction to the Poincar\'{e} subgroup \cite{Mack:1969dg}. Therefore the deformation parameter $\zeta$ can be identified with twice the helicity $h$ of the corresponding massless representation of the Poincar\'e group.
We  also extended  these results to the minimal unitary representations of supergroups $SU(2,2\,|\,N)$ with the even subgroup $SU(2,2)\times U(N)$  and their deformations.  The minimal unitary supermultiplet of $SU(2,2\,|\,N)$ coincides with the CPT self-conjugate (scalar) doubleton supermultiplet, and for $PSU(2,2\,|\,4)$ it is simply the four dimensional $N=4$  Yang-Mills supermultiplet.
 We showed that there exists a one-parameter family of deformations of the minimal unitary supermultiplet of $SU(2,2\,|\,N)$, and  each integer value of the  deformation parameter
$\zeta$ leads to  a unique unitary supermultiplet of $SU(2,2\,|\,N)$.    The minimal unitary supermultiplet of $SU(2,2\,|\,N)$ and its deformations coincide with  the unitary doubleton supermultiplets that were constructed and studied using the oscillator method earlier \cite{Gunaydin:1984fk,Gunaydin:1998sw,Gunaydin:1998jc}.  These  results extend to the minreps of $SU(m,n)$ and of $SU(m,n\,|N)$ and their deformations in a straightforward manner.

More recently we gave a detailed study of the minimal unitary representation (minrep) of $SO(6,2) \simeq SO^*(8)$
over an Hilbert space of functions of five variables, obtained by quantizing
its quasiconformal realization, and its deformations, and we constructed the minimal unitary supermultiplet of $OSp(8^*|2N)$ \cite{Fernando:2010dp}.
We showed  that there exists a family of
``deformations'' of the minrep of $SO^*(8)$ labeled by the spin $t$ of an
$SU(2)_T$ subgroup of the little group $SO(4)$ of lightlike vectors. These
deformed minreps  labeled by $t$ are positive energy unitary irreducible
representations of $SO^*(8)$ that describe massless conformal fields in six
dimensions. The $SU(2)_T$ spin $t$ is the six dimensional analog of $U(1)$ deformations of the minrep of $4D$ conformal group $SU(2,2)$ labeled by helicity. The minimal unitary
representation  of $OSp(8^*|2N)$  describes a massless six dimensional
conformal supermultiplet.  In particular, the minimal unitary supermultiplet of $OSp(8^*|4)$
is the massless supermultiplet of $(2,0)$ conformal field theory that is
believed to be dual to M-theory on $AdS_7 \times S^4$ . It is simply the scalar doubleton supermultiplet of  $OSp(8^*|4)$ first constructed in \cite{Gunaydin:1984wc}.

The oscillator construction of the positive energy unitary supermultiplets of $OSp(8^*|2N)$ was first given in \cite{Gunaydin:1984wc}. The  unitary supermultiplets of $OSp(8^*|2N)$ and their applications to $AdS_7/CFT_6$ dualities were further studied in \cite{Gunaydin:1999ci,Fernando:2001ak}, where it was shown that the doubleton supermultiplets correspond to massless conformal supermultiplets in six dimensions.
Construction of positive  energy unitary supermultiplets of $OSp(8^*|2N)$   using harmonic superspace methods as well as  their applications  to $AdS_7/CFT_6$ dualities were studied in  \cite{Ferrara:2000xg,Ferrara:2000dv}.
 A classification of  positive energy unitary supermultiplets of $6D$ superconformal algebras using Cartan-Kac formalism  was given in \cite{Minwalla:1997ka,Dobrev:2002dt}. The oscillator construction of positive energy unitary representations of general supergroups $OSp(2M^*|2N) $ with even subgroups $SO^*(2M)\times USp(2N)$ was given much earlier in \cite{Gunaydin:1990ag}.

In this paper we extend the results of \cite{Fernando:2010dp} and show that the minimal unitary supermultiplet of $OSp(8^*|2N)$ admits deformations labeled uniquely again by the spin $\mathfrak{t}$ of an $SU(2)_{\mathcal{T}}$ subgroup and   construct all  such deformed minimal unitary supermultiplets.
In section \ref{SO*(8)D}, we review our results on the minimal unitary representation of $SO^*(8)$ and its deformations realized over the Hilbert space of functions of five variables. In particular we give  a ``particle basis'' for these unitary representations over the tensor product of Fock space of four bosonic oscillators with the state space of a conformal (singular) oscillator. Their transformations under a distinguished $ SO(4)\times U(1)\times U(1)$ subgroup are also given.
In section \ref{OSp(8*|2N)D}, we present the deformations of the minimal unitary representation of
$OSp(8^*|2N)$ labeled by the spin $\mathfrak{t}$ of an $SU(2)_{\mathcal{T}}$ subgroup. Section \ref{3GrOSp(8*|2N)D} presents the compact 3-graded decomposition of the Lie superalgebra of $OSp(8^*|2N)$ with respect to the subsuperalgebra of $U(4|N)$.
In section \ref{deformedminrepsupermultiplets}, we give the general deformed minimal unitary representations of $OSp(8^*|2N)$ as $6D$ massless conformal supermultiplets. In section \ref{deformedminrepsupermultipletsN=2}, we study the deformed minimal unitary supermultiplets of $OSp(8^*|4)$ which is the symmetry superalgebra of eleven dimensional supergravity compactified over $AdS_7 \times S^4$. We show that the minimal unitary supermultiplet of  $OSp(8^*|4)$ and its deformations are precisely the doubleton supermultiplets that were constructed and studied using the twistorial oscillator construction \cite{Gunaydin:1984wc,Gunaydin:1999ci,Fernando:2001ak}. Appendix \ref{USp(2N)} reviews the construction of relevant representations of $USp(2N)$ using ``supersymmetry fermions.''


\section{Deformations of the Minimal Unitary Representation of $SO^*(8)$}
\label{SO*(8)D}

\renewcommand{\theequation}{\arabic{section}.\arabic{equation}}
\setcounter{equation}{0}

In our previous work \cite{Fernando:2010dp}, we gave a detailed study of  the minimal unitary
representation of $AdS_7$ or $Conf_6$ group $SO^*(8) \simeq SO(6,2)$ obtained by
the quantization of its quasiconformal realization \cite{Gunaydin:2006vz}. This minrep coincides
with the scalar doubleton representation of $SO^*(8)$, which corresponds to a
massless conformal scalar field in six dimensions \cite{Gunaydin:1984wc,Gunaydin:1999ci,Fernando:2001ak}.
There are infinitely many other doubleton representations of $SO^*(8)$,
corresponding to $6D$ massless conformal fields of higher spin
\cite{Gunaydin:1984wc,Gunaydin:1999ci,Fernando:2001ak}. In the oscillator
approach \cite{Gunaydin:1981dc,Gunaydin:1981yq}, all the doubleton
representations can be constructed over the Fock space of two pairs of
twistorial oscillators transforming in the spinor representation of $SO^*(8)$
\cite{Gunaydin:1984wc,Gunaydin:1999ci}. In \cite{Fernando:2010dp}, we obtained
 all these higher spin doubleton representations  from the
minimal unitary representation via a ``deformation'' in a manner similar to
what happens in the case of $4D$ conformal group $SU(2,2)$
\cite{Fernando:2009fq}.

In this section, we shall review how one  deforms the minimal unitary
representation of $SO^*(8)$ so as to obtain infinitely many irreducible
unitary representations that are isomorphic to  the irreducible doubleton representations of $SO^*(8)$.


\subsection{The 5-grading of the deformed minrep of $SO^*(8)$ with respect to
the subgroup $SO^*(4) \times SU(2) \times SO(1,1)$}
\label{5GrSO*(8)D}

The noncompact Lie algebra $\mathfrak{so}^*(8)$ has a 5-grading with respect
to its subalgebra $\mathfrak{g}^{(0)} = \mathfrak{so}^*(4) \oplus \mathfrak{su}(2) \oplus
\mathfrak{so}(1,1)$\cite{Gunaydin:2006vz}:
\begin{equation}
\mathfrak{so}^*(8)
= \mathfrak{g}^{(-2)} \oplus
  \mathfrak{g}^{(-1)} \oplus
  \left[
   \mathfrak{so}^*(4) \oplus \mathfrak{su}(2) \oplus \Delta
  \right] \oplus
  \mathfrak{g}^{(+1)} \oplus
  \mathfrak{g}^{(+2)}
\label{SO*(8)5Gr}
\end{equation}
such that
\begin{equation}
\commute{\Delta}{\mathfrak{g}^{(m)}} = m \, \mathfrak{g}^{(m)}
\end{equation}
where  $\Delta$ is the $SO(1,1)$ generator.\footnote{ We use the standard convention of denoting the groups with capital letters and the corresponding Lie algebras with small case letters.}
In this decomposition, the subspaces $\mathfrak{g}^{(\pm 2)}$ are
one-dimensional, and the subspaces $\mathfrak{g}^{(\pm 1)}$ transform in the
$\left( \mathbf{4} , \mathbf{2} \right)$ dimensional representation of
$SO^*(4) \times SU(2)$. Since $\mathfrak{so}^*(4) = \mathfrak{su}(1,1) \oplus
\mathfrak{su}(2)$, the grade zero subalgebra can be written as
\begin{equation}
\mathfrak{g}^{(0)}
= \mathfrak{su}(1,1)_N \oplus \mathfrak{su}(2)_A \oplus \mathfrak{su}(2)_T \oplus
  \mathfrak{so}(1,1)
\end{equation}
where we denoted the $\mathfrak{su}(1,1)$ and $\mathfrak{su}(2)$ subalgebras
of $\mathfrak{so}^*(4)$ as $\mathfrak{su}(1,1)_N$ and $\mathfrak{su}(2)_A$,
respectively, and the $\mathfrak{su}(2)$ that commutes with
$\mathfrak{so}^*(4)$ in equation (\ref{SO*(8)5Gr}) as $\mathfrak{su}(2)_T$.
In the undeformed case, this $\mathfrak{su}(2)$ was denoted as
$\mathfrak{su}(2)_S$ in \cite{Fernando:2010dp}. In the deformation of the
minimal unitary representation of $SO^*(8)$, the subalgebra
$\mathfrak{su}(2)_S$ gets extended to  the diagonal subalgebra
\eq
\mathfrak{su}(2)_T \subset \mathfrak{su}(2)_S \oplus \mathfrak{su}(2)_G
\en
 where the generators of
$\mathfrak{su}(2)_S$ are realized as bilinears of bosonic oscillators and those of
of $\mathfrak{su}(2)_G$ are realized in terms of
fermionic oscillators.\footnote{We should note  that in our previous
paper \cite{Fernando:2010dp}, we added a ``$\circ$'' above all deformed
generators to distinguish them from the undeformed generators. In this paper,
we drop those circles for the sake of simplicity.}

To realize the  minrep of $SO^*(8)$ and its deformations, one first  introduces bosonic annihilation operators
$a_m$, $b_m$ and their hermitian conjugates $a^m = \left( a_m \right)^\dag$,
$b^m = \left( b_m \right)^\dag$ ($m,n,\dots = 1,2$) that satisfy the
commutation relations:
\begin{equation}
\commute{a_m}{a^n}
= \commute{b_m}{b^n}
= \delta^n_m
\qquad \qquad
\commute{a_m}{a_n}
= \commute{a_m}{b_n}
= \commute{b_m}{b_n}
= 0
\end{equation}
and a single ``central-charge coordinate'' $x$ and its conjugate momentum $p$
such that
\begin{equation}
\commute{x}{p} = i \,.
\end{equation}
Now the generators of $\mathfrak{su}(2)_S$  are
realized as follows:
\begin{equation}
S_+ = a^m b_m
\qquad \qquad
S_- = \left( S_+ \right)^\dag
    = a_m b^m
\qquad \qquad
S_0 = \frac{1}{2} \left( N_a - N_b \right)
\label{SU(2)S_gen}
\end{equation}
where $N_a = a^m a_m$ and $N_b = b^m b_m$ are the respective number
operators. They satisfy:
\begin{equation}
\commute{S_+}{S_-} = 2 \, S_0
\qquad \qquad \qquad
\commute{S_0}{S_\pm} = \pm S_\pm
\end{equation}
The quadratic Casimir of $\mathfrak{su}(2)_S$ is
\begin{equation}
\begin{split}
\mathcal{C}_2 \left[ \mathfrak{su}(2)_S \right]
 = S^2
&= {S_0}^2 + \frac{1}{2} \left( S_+ S_- + S_- S_+ \right)
\\
&= \frac{1}{2} \left(N_a + N_b \right)
   \left[ \frac{1}{2} \left( N_a + N_b \right) + 1 \right]
   - 2 a^{[m} b^{n]} \, a_{[m} b_{n]}
\end{split}
\end{equation}
where square bracketing $a_{[m} b_{n]} = \frac{1}{2} \left( a_m b_n - a_n b_m
\right)$ represents antisymmetrization of weight one.

To realize $SU(2)_G$, we introduce an arbitrary number $P$ pairs of fermionic
annihilation operators $\xi_x$ and $\chi_x$ and their hermitian conjugates $\xi^x =
\left( \xi_x \right)^\dag$ and $\chi^x = \left( \chi_x \right)^\dag$ ($x =
1,2,\dots,P$) that satisfy the usual anti-commutation relations:
\begin{equation}
\anticommute{\xi_x}{\xi^y}
= \anticommute{\chi_x}{\chi^y}
= \delta^x_y
\qquad \qquad
\anticommute{\xi_x}{\xi_y}
= \anticommute{\xi_x}{\chi_y}
= \anticommute{\chi_x}{\chi_y}
= 0
\label{Dfermions}
\end{equation}
The generators of $SU(2)_G$ are given by the following bilinears of these
fermionic oscillators:
\begin{equation}
G_+ = \xi^x \chi_x
\qquad \qquad
G_- = \chi^x \xi_x
\qquad \qquad
G_0 = \frac{1}{2} \left( N_\xi - N_\chi \right)
\label{SU(2)G_gen}
\end{equation}
where $N_\xi = \xi^x \xi_x$ and $N_\chi = \chi^x \chi_x$ are the respective
number operators. They satisfy the commutation relations:\footnote{We should note that $SU(2)_G$, as defined in equation (\ref{SU(2)G_gen}),
commutes with the $USp(2P)$ group generated by the bilinears
$\xi_{(x} \chi_{y)}$, $\left( \xi^x \xi_y - \chi_y \chi^x \right)$ and 
$\xi^{(x} \chi^{y)}$.}
\begin{equation}
\commute{G_+}{G_-} = 2 \, G_0
\qquad \qquad \qquad
\commute{G_0}{G_\pm} = \pm G_0
\end{equation}
Then  the generators of $\mathfrak{su}(2)_T$ are simply:
\begin{equation}
\begin{split}
T_+
&= S_+ + G_+
 = a^m b_m + \xi^x \chi_x
\\
T_-
&= S_- + G_-
 = b^m a_m + \chi^x \xi_x
\\
T_0
&= S_0 + G_0
 = \frac{1}{2} \left( N_a - N_b + N_\xi - N_\chi \right)
\end{split}
\label{SU(2)T_gen}
\end{equation}
The $\mathfrak{su}(2)_G$ components realized in terms of fermions  represent the
deformations of the minrep. The quadratic Casimir of the subalgebra $\mathfrak{su}(2)_T$
is
\begin{equation}
\mathcal{C}_2 \left[ \mathfrak{su}(2)_T \right]
= T^2
= {T_0}^2 + \frac{1}{2} \left( T_+ T_- + T_- T_+ \right) \,.
\end{equation}

The generators of $\mathfrak{su}(2)_A$ and $\mathfrak{su}(1,1)_N$, which we denote as  $A_{\pm,0}$ and $N_{\pm,0}$, respectively, are realized purely in terms of bosonic  oscillators:
\begin{equation}
\begin{aligned}
A_+ &= a^1 a_2 + b^1 b_2
\\
A_- &= \left( A_+ \right)^\dag = a_1 a^2 + b_1 b^2
\\
A_0 &= \frac{1}{2} \left( a^1 a_1 - a^2 a_2 + b^1 b_1 - b^2 b_2 \right)
\end{aligned}
\qquad  \qquad
\begin{aligned}
N_+ &= a^1 b^2 - a^2 b^1
\\
N_- &= \left( N_+ \right)^\dag = a_1 b_2 - a_2 b_1
\\
N_0 &= \frac{1}{2} \left( N_a + N_b \right) + 1
\end{aligned}
\label{SU(2)AN_gen}
\end{equation}
and they do not get modified by $\xi$- and $\chi$-type fermionic oscillators under deformation.
They satisfy the commutation relations:
\begin{equation}
\begin{aligned}
\commute{A_+}{A_-} &= 2 \, A_0
\\
\commute{A_0}{A_\pm} &= \pm A_\pm
\end{aligned}
\qquad \qquad \qquad
\begin{aligned}
\commute{N_-}{N_+} &= 2 \, N_0
\\
\commute{N_0}{N_\pm} &= \pm N_\pm
\end{aligned}
\end{equation}
The quadratic Casimirs of these subalgebras
\begin{equation}
\begin{split}
\mathcal{C}_2 \left[ \mathfrak{su}(2)_A \right]
= A^2 &= {A_0}^2 + \frac{1}{2} \left( A_+ A_- + A_- A_+ \right)
\\
\mathcal{C}_2 \left[ \mathfrak{su}(1,1)_N \right]
= N^2 &= {N_0}^2 - \frac{1}{2} \left( N_+ N_- + N_- N_+ \right)
\end{split}
\end{equation}
coincide and are equal to that of $\mathfrak{su}(2)_S$ in the minrep:
\begin{equation}
S^2 = A^2 = N^2
\end{equation}

The generator $\Delta$ that defines the 5-grading is realized in terms of the
``central charge coordinate'' $x$ and its conjugate momentum $p$ as
\begin{equation}
\Delta = \frac{1}{2} \left( x p + p x \right) \,.
\end{equation}
The  generator in grade $-2$ space is given by
\begin{equation}
K_-
= \frac{1}{2} x^2
\end{equation}
and the eight generators in grade $-1$ subspace take the form:
\begin{equation}
\begin{aligned}
U_m &= x \, a_m
\\
V_m &= x \, b_m
\end{aligned}
\qquad \qquad \qquad \qquad
\begin{aligned}
U^m &= x \, a^m
\\
V^m &= x \, b^m
\end{aligned}
\end{equation}
Together with $K_-$, they form an Heisenberg algebra:
\begin{equation}
\begin{split}
\commute{U_m}{U^n}
&= \commute{V_m}{V^n}
 = 2 \, \delta^n_m \, K_-
\\
\commute{U_m}{U_n}
&= \commute{U_m}{V_n}
 = \commute{V_m}{V_n}
 = 0
\end{split}
\end{equation}

The single generator in grade $+2$ subspace is realized as follows:
\begin{equation}
K_+
= \frac{1}{2} p^2
  + \frac{1}{4 \, x^2} \left( 8 \, T^2 + \frac{3}{2} \right)
\end{equation}
The generators  $\Delta$, $K_{\pm}$ form a distinguished  $\mathfrak{su}(1,1)$ subalgebra,
that we  denote as $\mathfrak{su}(1,1)_K$:
\begin{equation}
\commute{K_-}{K_+} = i \, \Delta
\qquad \qquad \qquad
\commute{\Delta}{K_\pm} = \pm 2 i \, K_\pm
\end{equation}
Its quadratic Casimir operator turns out to be  equal to that of
$\mathfrak{su}(2)_T$:
\begin{equation}
\mathcal{C}_2 \left[ \mathfrak{su}(1,1)_K \right]
= K^2 = \frac{1}{2} (K_+ K_- + K_- K_+) - \frac{1}{4} \Delta^2
      =  T^2
\end{equation}

The generators in grade $+1$ subspace can be obtained by taking the
commutators of the form $\commute{\mathfrak{g}^{(-1)}}{\mathfrak{g}^{(+2)}}$:
\begin{equation}
\begin{split}
\widetilde{U}_m = i \commute{U_m}{K_+}
& \qquad \qquad \qquad \qquad
\widetilde{U}^m = \left( \widetilde{U}_m \right)^\dag
                = i \commute{U^m}{K_+}
\\
\widetilde{V}_m = i \commute{V_m}{K_+}
& \qquad \qquad \qquad \qquad
\widetilde{V}^m = \left( \widetilde{V}_m \right)^\dag
                = i \commute{V^m}{K_+}
\end{split}
\end{equation}
Explicitly they are given by:
\begin{equation}
\begin{split}
\widetilde{U}_m
&= - p \, a_m
   + \frac{2i}{x}
      \left[ \left( T_0 + \frac{3}{4} \right) a_m
             + T_- \, b_m
      \right]
\\
\widetilde{U}^m
&= - p \, a^m
   - \frac{2i}{x}
      \left[ \left( T_0 - \frac{3}{4} \right) a^m
             + T_+ \, b^m
      \right]
\\
\widetilde{V}_m
&= - p \, b_m
   - \frac{2i}{x}
      \left[ \left( T_0 - \frac{3}{4} \right) b_m
             - T_+ \, a_m
      \right]
\\
\widetilde{V}^m
&= - p \, b^m
   + \frac{2i}{x}
      \left[ \left( T_0 + \frac{3}{4} \right) b^m
             - T_- \, a^m
      \right]
\end{split}
\label{g+1bosonic}
\end{equation}
They form an Heisenberg algebra with $K_+$ as its ``central charge'':
\begin{equation}
\begin{split}
\commute{\widetilde{U}_m}{\widetilde{U}^n}
&= \commute{\widetilde{V}_m}{\widetilde{V}^n}
 = 2 \, \delta^n_m \, K_+
\\
\commute{\widetilde{U}_m}{\widetilde{U}_n}
&= \commute{\widetilde{U}_m}{\widetilde{V}_n}
 = \commute{\widetilde{V}_m}{\widetilde{V}_n}
 = 0
\end{split}
\end{equation}

The commutators $\commute{\mathfrak{g}^{(-2)}}{\mathfrak{g}^{(+1)}}$ close into grade $-1$ generators:
\begin{equation}
\begin{aligned}
\commute{\widetilde{U}_m}{K_-} &= i \, U_m
\\
\commute{\widetilde{V}_m}{K_-} &= i \, V_m
\end{aligned}
\qquad \qquad \qquad \qquad
\begin{aligned}
\commute{\widetilde{U}^m}{K_-} &= i \, U^m
\\
\commute{\widetilde{V}^m}{K_-} &= i \, V^m
\end{aligned}
\end{equation}

The non-vanishing commutators of the form
$\commute{\mathfrak{g}^{(-1)}}{\mathfrak{g}^{(+1)}}$ are:
\begin{equation}
\begin{split}
\commute{U_m}{\widetilde{U}^n}
&= - \delta^n_m \, \Delta
   - 2 i \, \delta^n_m \, N_0
   - 2 i \, \delta^n_m \, T_0
   - 2 i \, A^n_{~m}
\\
\commute{V_m}{\widetilde{V}^n}
&= - \delta^n_m \, \Delta
   - 2 i \, \delta^n_m \, N_0
   + 2 i \, \delta^n_m \, T_0
   - 2 i \, A^n_{~m}
\\
\commute{U_m}{\widetilde{V}^n}
&= - 2 i \, \delta^n_m \, T_-
\qquad \qquad
\commute{V_m}{\widetilde{U}^n}
 = - 2 i \, \delta^n_m \, T_+
\\
\commute{U_m}{\widetilde{V}_n}
&= - 2 i \, \epsilon_{mn} \, N_-
\qquad \qquad
\commute{V_m}{\widetilde{U}_n}
 = + 2 i \, \epsilon_{mn} \, N_-
\end{split}
\end{equation}
where $\epsilon_{mn}$ is the Levi-Civita tensor ($\epsilon_{12} = +1$) and we
have labeled the generators of $\mathfrak{su}(2)_A$ as $A^m_{~n}$:
\begin{equation}
A^1_{~1} = - A^2_{~2} = A_0
\qquad \qquad \qquad
A^1_{~2} = A_+
\qquad \qquad \qquad
A^2_{~1} = \left( {A^1_{~2}} \right)^\dag = A_-
\end{equation}
With the generators defined above, the 5-graded decomposition of the deformed
minimal unitary realization, which we denote as $\mathfrak{so}^*(8)_D$,
takes the form:
\begin{equation}
\begin{split}
\mathfrak{so}^*(8)_D
&= \mathfrak{g}^{(-2)}_D \oplus
   \mathfrak{g}^{(-1)}_D \oplus
   \left[
    \mathfrak{so}^*(4) \oplus \mathfrak{su}(2)_T \oplus \Delta
   \right] \oplus
   \mathfrak{g}^{(+1)}_D \oplus
   \mathfrak{g}^{(+2)}_D
\\
&= ~ \mathbf{1} ~~ \oplus
   ~~ \left( \mathbf{4} , \mathbf{2} \right) ~ \oplus
   \left[ \mathfrak{su}(2)_A \oplus
          \mathfrak{su}(1,1)_N \oplus
          \mathfrak{su}(2)_T \oplus
          \mathfrak{so}(1,1)_{\Delta}
   \right] \oplus
   ~ \left( \mathbf{4} , \mathbf{2} \right) ~ \oplus
   ~ \mathbf{1}
\\
&= K_-
   \oplus
   \left[ U_m \,,\, U^m \,,\, V_m \,,\, V^m \right]
   \oplus
   \left[ ~ A_{\pm,0} ~ \oplus ~ N_{\pm,0} ~ \oplus ~ T_{\pm,0} ~ \oplus
          ~ \Delta ~ \right]
\\
& \qquad \qquad \qquad \qquad \qquad \qquad \qquad \qquad \qquad \qquad \quad
   \oplus
   \left[ \widetilde{U}_m \,,\, \widetilde{U}^m \,,\,
          \widetilde{V}_m \,,\, \widetilde{V}^m
   \right]
   \oplus
   K_+ \,
\end{split}
\label{SO*(8)D5Gr}
\end{equation}

The quadratic Casimir of $\mathfrak{so}^*(8)_D$ is given by
\begin{equation}
\begin{split}
\mathcal{C}_2 \left[ \mathfrak{so}^*(8)_D \right]
&= \mathcal{C}_2 \left[ \mathfrak{su}(2)_T \right]
   + \mathcal{C}_2 \left[ \mathfrak{su}(2)_A \right]
   + \mathcal{C}_2 \left[ \mathfrak{su}(1,1)_N \right]
   + \mathcal{C}_2 \left[ \mathfrak{su}(1,1)_K \right]
\\
& \quad
   - \frac{i}{4} \,
    \mathcal{F} \left( U , \widetilde{U} , V , \widetilde{V}\right)
\end{split}
\end{equation}
where
\begin{equation}
\begin{split}
\mathcal{F} \left( U , \widetilde{U} , V , \widetilde{V}\right)
&= \left( U_m \widetilde{U}^m + V_m \widetilde{V}^m
          + \widetilde{U}^m U_m + \widetilde{V}^m V_m \right)
\\
& \qquad
   - \left( U^m \widetilde{U}_m + V^m \widetilde{V}_m
         + \widetilde{U}_m U^m + \widetilde{V}_m V^m \right)
\end{split}
\end{equation}
and reduces to
\begin{equation}
\mathcal{C}_2 \left[ \mathfrak{so}^*(8)_D \right]
= 2 \, G^2 - 4
\label{QC_SO*(8)D}
\end{equation}
 where $G^2$ is the quadratic Casimir of
$\mathfrak{su}(2)_G$. Thus the quadratic Casimir of the deformed minrep of $SO^*(8)$ depends only on the quadratic Casimir of $SU(2)_G$ constructed out of $\xi$- and $\chi$-type fermionic oscillators used to deform the minrep.


\subsection{The noncompact 3-grading of $SO^*(8)_D$ with respect to the
subgroup $SU^*(4) \times SO(1,1)$}
\label{NC3GrSO*(8)D}

Considered as the six dimensional conformal group, $SO^*(8)_D$ has a
noncompact 3-grading determined by the dilatation generator $\mathcal{D}$:
\begin{equation}
\mathfrak{so}^*(8)_D
= \mathfrak{N}^-_D \oplus
  \mathfrak{N}^0_D \oplus
  \mathfrak{N}^+_D
\end{equation}
where $\mathfrak{N}^0_D = \mathfrak{su}^*(4) \oplus \mathfrak{so}(1,1)$, with $\mathfrak{su}^*(4) \simeq \mathfrak{so}(5,1)$ corresponding to the six dimensional Lorentz group. The $\mathfrak{so}(1,1)$ dilatation
generator is given by
\begin{equation}
\mathcal{D}
= \frac{1}{2} \left[ \Delta - i \left( N_+ - N_- \right) \right] \,.
\end{equation}
The generators that belong to $\mathfrak{N}^{\pm}_D$ and $\mathfrak{N}^0_D$
subspaces are as follows:
\begin{equation}
\begin{split}
\mathfrak{N}^-_D
&= K_-
   \oplus \left[ N_0 - \frac{1}{2} \left( N_+ + N_- \right) \right]
\\
& \quad
   \oplus \left( U^1 - V_2 \right)
   \oplus \left( U^2 + V_1 \right)
   \oplus \left( V^1 + U_2 \right)
   \oplus \left( V^2 - U_1 \right)
\\
\mathfrak{N}^0_D
&= \mathcal{D}
   \oplus \frac{1}{2} \left[ \Delta + i \left( N_+ - N_- \right) \right]
   \oplus T_{\pm,0}
   \oplus A_{\pm,0}
\\
& \quad
   \oplus \left( U^1 + V_2 \right)
   \oplus \left( U^2 - V_1 \right)
   \oplus \left( V^1 - U_2 \right)
   \oplus \left( V^2 + U_1 \right)
\\
& \quad
   \oplus \left( \widetilde{U}^1 - \widetilde{V}_2 \right)
   \oplus \left( \widetilde{U}^2 + \widetilde{V}_1 \right)
   \oplus \left( \widetilde{V}^1 + \widetilde{U}_2 \right)
   \oplus \left( \widetilde{V}^2 - \widetilde{U}_1 \right)
\\
\mathfrak{N}^+_D
&= K_+
   \oplus \left[ N_0 + \frac{1}{2} \left( N_+ + N_- \right) \right]
\\
& \quad
   \oplus \left( \widetilde{U}^1 + \widetilde{V}_2 \right)
   \oplus \left( \widetilde{U}^2 - \widetilde{V}_1 \right)
   \oplus \left( \widetilde{V}^1 - \widetilde{U}_2 \right)
   \oplus \left( \widetilde{V}^2 + \widetilde{U}_1 \right)
\end{split}
\end{equation}
 In terms of the above operators, the
Lorentz group generators $\mathcal{M}_{\mu\nu}$ ($\mu,\nu,\dots = 0,1,2,\dots,5$)
in six dimensions belonging to $\mathfrak{N}^0_D $ are given by:
\begin{subequations}
\begin{equation}
\begin{aligned}
\mathcal{M}_{01}
&= \frac{1}{4} \left[ \left( U^1 + V_2 \right)
                      + \left( V^2 + U_1 \right)
                      + i \left( \widetilde{U}^1 - \widetilde{V}_2 \right)
                      + i \left( \widetilde{V}^2 - \widetilde{U}_1 \right)
               \right]
\\
\mathcal{M}_{02}
&= \frac{i}{4} \left[ \left( U^1 + V_2 \right)
                      - \left( V^2 + U_1 \right)
                      + i \left( \widetilde{U}^1 - \widetilde{V}_2 \right)
                      - i \left( \widetilde{V}^2 - \widetilde{U}_1 \right)
               \right]
\\
\mathcal{M}_{03}
&= \frac{i}{4} \left[ \left( U^2 - V_1 \right)
                      + \left( V^1 - U_2 \right)
                      + i \left( \widetilde{U}^2 + \widetilde{V}_1 \right)
                      + i \left( \widetilde{V}^1 + \widetilde{U}_2 \right)
               \right]
\\
\mathcal{M}_{04}
&= - \frac{1}{4} \left[ \left( U^2 - V_1 \right)
                        - \left( V^1 - U_2 \right)
                        + i \left( \widetilde{U}^2 + \widetilde{V}_1 \right)
                        - i \left( \widetilde{V}^1 + \widetilde{U}_2 \right)
                 \right]
\end{aligned}
\end{equation}
\begin{equation}
\begin{aligned}
\mathcal{M}_{15}
&= \frac{1}{4} \left[ \left( U^1 + V_2 \right)
                      + \left( V^2 + U_1 \right)
                      - i \left( \widetilde{U}^1 - \widetilde{V}_2 \right)
                      - i \left( \widetilde{V}^2 - \widetilde{U}_1 \right)
               \right]
\\
\mathcal{M}_{25}
&= \frac{i}{4} \left[ \left( U^1 + V_2 \right)
                      - \left( V^2 + U_1 \right)
                      - i \left( \widetilde{U}^1 - \widetilde{V}_2 \right)
                      + i \left( \widetilde{V}^2 - \widetilde{U}_1 \right)
               \right]
\\
\mathcal{M}_{35}
&= \frac{i}{4} \left[ \left( U^2 - V_1 \right)
                      + \left( V^1 - U_2 \right)
                      - i \left( \widetilde{U}^2 + \widetilde{V}_1 \right)
                      - i \left( \widetilde{V}^1 + \widetilde{U}_2 \right)
               \right]
\\
\mathcal{M}_{45}
&= - \frac{1}{4} \left[ \left( U^2 - V_1 \right)
                        - \left( V^1 - U_2 \right)
                        - i \left( \widetilde{U}^2 + \widetilde{V}_1 \right)
                        + i \left( \widetilde{V}^1 + \widetilde{U}_2 \right)
                 \right]
\end{aligned}
\end{equation}
\begin{equation}
\begin{aligned}
\mathcal{M}_{12}
&= T_0 + A_0
\\
\mathcal{M}_{14}
&= \frac{i}{2} \left( T_+ - T_- - A_+ + A_- \right)
\\
\mathcal{M}_{24}
&= - \frac{1}{2} \left( T_+ + T_- - A_+ - A_- \right)
\end{aligned}
\qquad \qquad
\begin{aligned}
\mathcal{M}_{13}
&= \frac{1}{2} \left( T_+ + T_- + A_+ + A_- \right)
\\
\mathcal{M}_{23}
&= \frac{i}{2} \left( T_+ - T_- + A_+ - A_- \right)
\\
\mathcal{M}_{34}
&= T_0 - A_0
\end{aligned}
\end{equation}
\begin{equation}
\mathcal{M}_{05}
= \frac{1}{2} \left[ \Delta + i \left( N_+ - N_- \right) \right]
\end{equation}
\end{subequations}
They  satisfy the commutation
relations
\begin{equation}
\commute{\mathcal{M}_{\mu\nu}}{\mathcal{M}_{\rho\tau}}
= i \left(
     \eta_{\nu\rho} \mathcal{M}_{\mu\tau}
     - \eta_{\mu\rho} \mathcal{M}_{\nu\tau}
     - \eta_{\nu\tau} \mathcal{M}_{\mu\rho}
     + \eta_{\mu\tau} \mathcal{M}_{\nu\rho}
    \right)
\end{equation}
where $\eta_{\mu\nu} = \mathrm{diag} (-,+,+,+,+,+)$.
The six generators of grade $+1$ space are the momenta
$\mathcal{P}_\mu$ that generate translations, and the six generators of grade $-1$ space are the special
conformal generators $\mathcal{K}_\mu$ ($\mu = 0,1,2,\dots,5$):
\begin{equation}
\begin{aligned}
\mathcal{P}_0
&= K_+ + \left[ N_0 + \frac{1}{2} \left( N_+ + N_- \right) \right]
\\
\mathcal{P}_1
&= - \frac{1}{2} \left[
                  \left( \widetilde{U}^1 + \widetilde{V}_2 \right)
                  + \left( \widetilde{V}^2 + \widetilde{U}_1 \right)
                 \right]
\\
\mathcal{P}_2
&= - \frac{i}{2} \left[
                  \left( \widetilde{U}^1 + \widetilde{V}_2 \right)
                  - \left( \widetilde{V}^2 + \widetilde{U}_1 \right)
                 \right]
\\
\mathcal{P}_3
&= - \frac{i}{2} \left[
                  \left( \widetilde{U}^2 - \widetilde{V}_1 \right)
                  + \left( \widetilde{V}^1 - \widetilde{U}_2 \right)
                 \right]
\\
\mathcal{P}_4
&= \frac{1}{2} \left[
                \left( \widetilde{U}^2 - \widetilde{V}_1 \right)
                - \left( \widetilde{V}^1 - \widetilde{U}_2 \right)
               \right]
\\
\mathcal{P}_5
&= K_+ - \left[ N_0 + \frac{1}{2} \left( N_+ + N_- \right) \right]
\end{aligned}
\qquad \qquad
\begin{aligned}
\mathcal{K}_0
&= \left[ N_0 - \frac{1}{2} \left( N_+ + N_- \right) \right] + K_-
\\
\mathcal{K}_1
&= \frac{i}{2} \left[
                \left( U^1 - V_2 \right) + \left( V^2 - U_1 \right)
               \right]
\\
\mathcal{K}_2
&= - \frac{1}{2} \left[
                  \left( U^1 - V_2 \right) - \left( V^2 - U_1 \right)
                 \right]
\\
\mathcal{K}_3
&= - \frac{1}{2} \left[
                  \left( U^2 + V_1 \right) + \left( V^1 + U_2 \right)
                 \right]
\\
\mathcal{K}_4
&= - \frac{i}{2} \left[
                  \left( U^2 + V_1 \right) - \left( V^1 + U_2 \right)
                 \right]
\\
\mathcal{K}_5
&= \left[ N_0 - \frac{1}{2} \left( N_+ + N_- \right) \right] - K_-
\end{aligned}
\label{SO*(8)D_PKgenerators}
\end{equation}
They
satisfy the commutation relations:
\begin{equation}
\begin{split}
\commute{\mathcal{D}}{\mathcal{P}_\mu}
&= + i \, \mathcal{P}_\mu
\qquad \qquad
\commute{\mathcal{D}}{\mathcal{K}_\mu}
 = - i \, \mathcal{K}_\mu
\\
\commute{\mathcal{D}}{\mathcal{M}_{\mu\nu}}
&= \commute{\mathcal{P}_\mu}{\mathcal{P}_\nu}
 = \commute{\mathcal{K}_\mu}{\mathcal{K}_\nu}
 = 0
\\
\commute{\mathcal{P}_\mu}{\mathcal{M}_{\nu\rho}}
&= i \left( \eta_{\mu\nu} \, \mathcal{P}_\rho
            - \eta_{\mu\rho} \, \mathcal{P}_\nu
     \right)
\\
\commute{\mathcal{K}_\mu}{\mathcal{M}_{\nu\rho}}
&= i \left( \eta_{\mu\nu} \, \mathcal{K}_\rho
            - \eta_{\mu\rho} \, \mathcal{K}_\nu
     \right)
\\
\commute{\mathcal{P}_\mu}{\mathcal{K}_\nu}
&= 2 i \left( \eta_{\mu\nu} \, \mathcal{D} + \mathcal{M}_{\mu\nu} \right)
\end{split}
\end{equation}

Interestingly, none of the special conformal transformations
$\mathcal{K}_\mu$ ($\mu = 0,1,2,\dots,5$) receives any contributions from
$\xi$- or $\chi$-type fermionic oscillators used to deform the minrep of $SO^*(8)$.
Furthermore,   the six dimensional Poincar\'e mass
operator vanishes identically:
\begin{equation}
\mathcal{M}^2 = \eta_{\mu\nu} \mathcal{P}^\mu \mathcal{P}^\nu = 0
\end{equation}
for the deformed minimal unitary realization of $SO^*(8)$ given above.
Hence each deformed irreducible minrep  corresponds to a massless conformal field in six
dimensions.


\subsection{The compact 3-grading of $SO^*(8)_D$ with respect to the subgroup
$SU(4) \times U(1)$}
\label{C3GrSO*(8)D}

The Lie algebra of $\mathfrak{so}^*(8)_D$ can be given a compact 3-grading
\begin{equation}
\mathfrak{so}^*(8)_D
= \mathfrak{C}^-_D \oplus \mathfrak{C}^0_D \oplus \mathfrak{C}^+_D
\end{equation}
with respect to its maximal compact subalgebra $\mathfrak{C}^0_D =
\mathfrak{su}(4) \oplus \mathfrak{u}(1)$, determined by the $\mathfrak{u}(1)$
generator:
\begin{equation}
H = N_0 + \frac{1}{2} \left( K_+ + K_- \right)
\label{BosonicHamiltonian}
\end{equation}
This $\mathfrak{u}(1)$ generator plays the role of the $AdS$ energy or the
conformal Hamiltonian when $SO^*(8) \simeq SO(6,2)$ is taken as the seven
dimensional $AdS$ group or the six dimensional conformal group, respectively.
In terms of the time components of momenta and special conformal generators defined in the  noncompact 3-graded basis (equation (\ref{SO*(8)D_PKgenerators})), we have
\begin{equation}
H = \frac{1}{2} \left( \mathcal{K}_0 +
\mathcal{P}_0 \right) \,.
\end{equation}

The  grade $-1$ operators in the compact basis are given by:
\begin{equation}
\begin{split}
Y_m &= \frac{1}{2} \left( U_m - i \, \widetilde{U}_m \right)
     = \frac{1}{2} \left( x + i \, p \right) a_m
       + \frac{1}{x}
         \left[
          \left( T_0 + \frac{3}{4} \right) a_m + T_- b_m
         \right]
\\
Z_m &= \frac{1}{2} \left( V_m - i \, \widetilde{V}_m \right)
     = \frac{1}{2} \left( x + i \, p \right) b_m
       - \frac{1}{x}
         \left[
          \left( T_0 - \frac{3}{4} \right) b_m - T_+ a_m
         \right]
\\
N_- &= a_1 b_2 - a_2 b_1
\\
B_- &= \frac{i}{2} \left[ \Delta + i \left( K_+ - K_- \right) \right]
     = \frac{1}{4} \left( x + i \, p \right)^2
       - \frac{1}{x^2}
          \left( T^2 + \frac{3}{16} \right)
\end{split}
\label{SO*(8)DGr-1}
\end{equation}
and the grade $+1$ operators are given by their hermitian conjugates:
\begin{equation}
\begin{split}
Y^m &= \frac{1}{2} \left( U^m + i \, \widetilde{U}^m \right)
     = \frac{1}{2} \left( x - i \, p \right) a^m
       + \frac{1}{x}
         \left[
          \left( T_0 - \frac{3}{4} \right) a^m + T_+ b^m
         \right]
\\
Z^m &= \frac{1}{2} \left( V^m + i \, \widetilde{V}^m \right)
     = \frac{1}{2} \left( x - i \, p \right) b^m
       - \frac{1}{x}
         \left[
          \left( T_0 + \frac{3}{4} \right) b^m - T_- a^m
         \right]
\\
N_+ &= a^1 b^2 - a^2 b^1
\\
B_+ &= - \frac{i}{2} \left[ \Delta - i \left( K_+ - K_- \right) \right]
     = \frac{1}{4} \left( x - i \, p \right)^2
       - \frac{1}{x^2}
          \left( T^2 + \frac{3}{16} \right)
\end{split}
\label{SO*(8)DGr+1}
\end{equation}
The $\mathfrak{su}(4)$ subalgebra has the maximal subalgebra
\begin{equation*}
\mathfrak{su}(4) \supset
\mathfrak{su}(2)_T \oplus \mathfrak{su}(2)_A \oplus \mathfrak{u}(1)_J
\end{equation*}
where the $\mathfrak{u}(1)_J$ generator is given by:
\begin{equation}
J = N_0 - \frac{1}{2} \left( K_+ + K_- \right)=\frac{1}{2} \left( \mathcal{K}_5 - \mathcal{P}_5 \right)
\end{equation}
The generators $T_{\pm,0}$ and $A_{\pm,0}$ of $\mathfrak{su}(2)_T$ and $\mathfrak{su}(2)_A$  were given in equations
(\ref{SU(2)T_gen}) and (\ref{SU(2)AN_gen}) and the generators belonging to the coset
\begin{equation*}
SU(4) \,/\, \left[ SU(2)_T \times SU(2)_A \times U(1)_J \right]
\end{equation*}
are as follows:
\begin{equation}
\begin{split}
C_{1m} &= \frac{1}{2} \left( U_m + i \, \widetilde{U}_m \right)
        = \frac{1}{2} \left( x - i \, p \right) a_m
          - \frac{1}{x}
            \left[ \left( T_0 + \frac{3}{4} \right) a_m + T_- b_m \right]
\\
C^{1m} &= \frac{1}{2} \left( U^m - i \, \widetilde{U}^m \right)
        = \frac{1}{2} \left( x + i \, p \right) a^m
          - \frac{1}{x}
            \left[ \left( T_0 - \frac{3}{4} \right) a^m + T_+ b^m \right]
\\
C_{2m} &= \frac{1}{2} \left( V_m + i \, \widetilde{V}_m \right)
        = \frac{1}{2} \left( x - i \, p \right) b_m
          + \frac{1}{x}
            \left[ \left( T_0 - \frac{3}{4} \right) b_m - T_+ a_m \right]
\\
C^{2m} &= \frac{1}{2} \left( V^m - i \, \widetilde{V}^m \right)
        = \frac{1}{2} \left( x + i \, p \right) b^m
          + \frac{1}{x}
            \left[ \left( T_0 + \frac{3}{4} \right) b^m - T_- a^m \right]
\end{split}
\label{SU(4)coset}
\end{equation}

The $\mathfrak{u}(1)$ generator $H$ that defines the compact 3-grading is an operator whose spectrum is bounded from below. Hence the minrep of $SO^*(8)$ and its deformations are all positive energy unitary (lowest weight) representations.
The unitary lowest weight representations of $SO^*(8)_D$ are uniquely
labeled by a lowest energy K-type, that transforms irreducibly under the
$SU(4)$ subgroup, with the lowest energy eigenvalue with respect to the $U(1)$
generator $H$, and are annihilated by all the grade $-1$ operators in
$\mathfrak{C}^-_D$.  Since $SU(2)_T \times SU(2)_A \times
U(1)_J$ is a maximal subgroup of $SU(4)$, one can label these lowest energy K-types
 by the  $SU(2)_T \times SU(2)_A \times U(1)_J$ quantum numbers of their highest weight vectors as irreps of $SU(4)$.


\subsection{Distinguished $SU(1,1)_K$ subgroup of $SO^*(8)_D$ generated by the
isotonic (singular) oscillators}
\label{SU(1,1)ofSO*(8)D}

Note that the $\mathfrak{u}(1)$ generator $H$, given in equation
(\ref{BosonicHamiltonian}), that determines the compact 3-grading of
$\mathfrak{so}^*(8)_D$ can be written as
\begin{equation}
H = H_a + H_b + H_\odot
\end{equation}
where
\begin{equation}
H_a = \frac{1}{2} \left( N_a + 2 \right)
\qquad \qquad \qquad
H_b = \frac{1}{2} \left( N_b + 2 \right)
\end{equation}
are simply  the Hamiltonians of standard bosonic  oscillators of $a$- and $b$-type.
On the other hand,
\begin{equation}
\begin{split}
H_\odot
 = \frac{1}{2} \left( K_+ + K_- \right)
&= \frac{1}{4} \left( x^2 + p^2 \right)
   + \frac{1}{x^2} \left( T^2 + \frac{3}{16} \right)
\\
&= \frac{1}{4} \left( x^2 - \frac{\partial^2}{\partial x^2} \right)
   + \frac{1}{x^2} \left( T^2 + \frac{3}{16} \right)
\end{split}
\label{SingularHamiltonian}
\end{equation}
is the Hamiltonian of a singular harmonic oscillator with a singular potential
function
\begin{equation}
V_D \left( x \right) = \frac{G_D}{x^2}
\qquad \mbox{where} \quad
G_D = 2 \, T^2 + \frac{3}{8} \,.
\end{equation}
$H_\odot$ also arises as the Hamiltonian of conformal quantum mechanics
\cite{de Alfaro:1976je} with $G_D$ playing the role of the coupling constant
\cite{Gunaydin:2001bt}. In some literature it is also referred to as the
isotonic oscillator \cite{Casahorran:1995vt,carinena-2007}.

Together with the generators $B_\pm$ belonging to  $\mathfrak{C}^\pm_D$ subspaces of
$\mathfrak{so}^*(8)_D$ (equations (\ref{SO*(8)DGr-1}) and
(\ref{SO*(8)DGr+1})):
\begin{equation}
\begin{split}
B_-
  = \frac{i}{2} \left[ \Delta + i \left( K_+ - K_- \right) \right]
 &= \frac{1}{4} \left( x + i p \right)^2
    - \frac{1}{x^2} \left( T^2 + \frac{3}{16} \right)
\\
 &= \frac{1}{4} \left( x + \frac{\partial}{\partial x} \right)^2
    - \frac{1}{x^2} \left( T^2 + \frac{3}{16} \right)
\\
B_+
  = - \frac{i}{2} \left[ \Delta - i \left( K_+ - K_- \right) \right]
 &= \frac{1}{4} \left( x - i p \right)^2
    - \frac{1}{x^2} \left( T^2 + \frac{3}{16} \right)
\\
 &= \frac{1}{4} \left( x - \frac{\partial}{\partial x} \right)^2
    - \frac{1}{x^2} \left( T^2 + \frac{3}{16} \right)
\end{split}
\end{equation}
$H_\odot$ generates the distinguished $\mathfrak{su}(1,1)_K$
subalgebra:\footnote{This is the  $\mathfrak{su}(1,1)$ subalgebra generated by the longest
root vector.}
\begin{equation}
\commute{B_-}{B_+} = 2 \, H_\odot
\qquad \qquad
\commute{H_\odot}{B_\pm} = \pm \, B_\pm
\end{equation}
For a given eigenvalue $t \left( t + 1 \right)$ of the quadratic Casimir
$T^2$ of $\mathfrak{su}(2)_T$, the wave functions corresponding to the
lowest energy eigenvalue of this singular harmonic oscillator Hamiltonian
will be superpositions of functions of the form
$\psi_0^{(\alpha_t)} \left( x \right) \Lambda \left( t , m_t \right)$, where
$\Lambda \left( t , m_t \right)$ is an eigenstate of $T^2$ and $T_0$, independent of $x$:
\begin{equation}
T^2 \, \Lambda \left( t , m_t \right)
= t \left( t + 1 \right) \Lambda \left( t , m_t \right)
\qquad \qquad \qquad
T_0 \, \Lambda \left( t , m_t \right)
= m_t \, \Lambda \left( t , m_t \right)
\end{equation}
and $\psi_0^{(\alpha_t)} \left( x \right)$ is a function of $x$ that satisfies
\begin{equation}
B_- \, \psi_0^{(\alpha_t)} \left( x \right) \Lambda \left( t , m_t \right)
= 0
\end{equation}
whose solution is given by \cite{MR858831}
\begin{equation}
\psi_0^{(\alpha_t)} \left( x \right)
= C_0 \, x^{\alpha_t} \, e^{-x^2/2}
\end{equation}
where $C_0$ is a normalization constant and
\begin{equation}
\alpha_t
= \frac{1}{2} + \sqrt{1 + 4 \, t \left( t + 1 \right)}
= 2 \, t + \frac{3}{2} \,.
\end{equation}
The normalizability of the state imposes the constraint
\begin{equation}
\alpha_t \geq \frac{1}{2} \,.
\end{equation}

A state of the form  $\psi_0^{(\alpha_t = 2 \, t + 3/2)} \left( x
\right) \Lambda \left( t , m_t \right)$ is an
eigenstate of $H_\odot$ with eigenvalue $\left( t + 1 \right)$:
\begin{equation}
H_\odot \, \psi_0^{(2 \, t + 3/2)} \left( x \right)
\Lambda \left( t , m_t \right)
= \left( t + 1 \right) \,
  \psi_0^{(2 \, t + 3/2)} \left( x \right)
  \Lambda \left( t , m_t \right)
\end{equation}
which is the lowest energy eigenvalue in the deformed case with the
deformation parameter $t$.
Higher energy eigenstates of $H_\odot$ can be obtained from
$\psi_0^{(2 \, t + 3/2)} \left( x \right) \Lambda \left( t , m_t \right)$ by acting on it repeatedly with the raising generator $B_+$:
\begin{equation}
\psi_n^{(2 \, t + 3/2)} \left( x \right)
\Lambda \left( t , m_t \right)
= C_n \, \left( B_+ \right)^n
  \psi_0^{(2 \, t + 3/2)} \left( x \right)
  \Lambda \left( t , m_t \right)
\label{isotonicgroundstate}
\end{equation}
where $C_n$ are normalization constants. They correspond to energy
eigenvalues $n + t + 1$:
\begin{equation}
H_\odot \, \psi_n^{(2 \, t + 3/2)} \left( x \right)
\Lambda \left( t , m_t \right)
= \left( n + t + 1 \right) \,
  \psi_n^{(2 \, t + 3/2)} \left( x \right)
  \Lambda \left( t , m_t \right)
\end{equation}
We shall denote the corresponding states as
\begin{equation*}
\ket{\psi_n^{(2 \, t + 3/2)} \left( x \right) \,;\,
\Lambda \left( t , m_t \right)}
= \ket{\psi_n^{(2 \, t + 3/2)} \left( x \right)}
  \otimes \ket{\Lambda \left( t , m_t \right)}
\end{equation*}
and refer to them as the particle basis of the state space of the (isotonic)
singular oscillator. The $\left( 2 \, t + 1 \right)$ states belonging
to the subspace corresponding to an irrep of $SU(2)_T$ labeled by spin
$t$ will all have the same eigenvalue of $H_\odot$.


\subsection{$SU(2)_T \times SU(2)_A \times U(1)_J\times U(1)_H$  basis of the deformed
minimal unitary representations of $SO^*(8)$ }
\label{SU2SU2U1}

The fermionic Fock vacuum $\ket{0}_F$ is chosen such that:
\begin{equation}
\xi_x \ket{0}_F = \chi_x \ket{0}_F = 0 \qquad \qquad x = 1,2,\dots,P
\end{equation}
A ``particle basis'' of states in the fermionic  Fock space is provided by the action
of creation operators $\xi^x$ and $ \chi^y$ on the Fock vacuum $\ket{0}_F$.
A state of the form\footnote{Note that square bracketing of fermionic
indices implies complete anti-symmetrization of weight one.}
\begin{equation*}
\chi^{[x_1} \chi^{x_2} \chi^{x_3} \dots \chi^{x_P]} \ket{0}_F
\end{equation*}
has a definite eigenvalue $-\frac{P}{2}$ of $G_0$ and is annihilated by the
lowering operator $G_-$. By repeatedly acting on this state with the
raising operator $G_+$, one can obtain $P$ other states of the form:
\begin{equation*}
\xi^{[x_1} \chi^{x_2} \chi^{x_3} \dots \chi^{x_P]} \ket{0}_F
\, \oplus \,
\xi^{[x_1} \xi^{x_2} \chi^{x_3} \dots \chi^{x_P]} \ket{0}_F
\, \oplus \,
\dots \dots
\, \oplus \,
\xi^{[x_1} \xi^{x_2} \xi^{x_3} \dots \xi^{x_P]} \ket{0}_F
\end{equation*}
We shall denote these $P+1$ states as
\begin{equation}
\ket{\frac{P}{2} \,,\, m_P} \qquad \qquad
\mbox{where} \quad
m_P = -\frac{P}{2} , -\frac{P}{2}+1 , \dots , +\frac{P}{2} \,.
\end{equation}
They  transform  irreducibly under $\mathfrak{su}(2)_G$ in
the spin $\frac{P}{2}$ representation.
We shall denote the bosonic Fock vacuum annihilated by all bosonic oscillators
$a_m , b_m$ ($m = 1,2$) as $\ket{0}_B$:
\begin{equation}
a_m \ket{0}_B = b_m \ket{0}_B=  0
\end{equation}
and the tensor product of fermionic and bosonic vacua simply as $\ket{0}$.

The tensor products of the states of the form $\left( a^m \right)^{n_{a,m}}
\ket{0}_B$, $\left( b^m  \right)^{n_{b,m}} \ket{0}_B$, $\xi^x \ket{0}_F$ and
$\chi^x \ket{0}_F$, where $n_{a,m}$ and $n_{b,m}$ are non-negative integers,
form the ``particle basis'' of states in the full  Fock space.
As the ``particle basis'' of the Hilbert space of the deformed minimal
unitary representation of $SO^*(8)$, we shall take the following tensor
products of the above states with the state space of the singular (isotonic) oscillator:
\begin{equation*}
\left( a^1 \right)^{n_{a,1}}
\left( a^2 \right)^{n_{a,2}}
\left( b^1 \right)^{n_{b,1}}
\left( b^2 \right)^{n_{b,2}} \ket{0}_B \otimes
\xi^{[x_1} \dots \xi^{x_k} \chi^{x_{k+1}} \dots \chi^{x_P]} \ket{0}_F \otimes
\ket{\psi_n^{(\alpha_t)}}
\end{equation*}
where square brackets imply full anti-symmetrization with weight one. We denote them as
\begin{equation*}
\left( a^1 \right)^{n_{a,1}} \left( a^2 \right)^{n_{a,2}}
\left( b^1 \right)^{n_{b,1}} \left( b^2 \right)^{n_{b,2}}
\xi^{[x_1} \dots \xi^{x_k} \chi^{x_{k+1}} \dots \chi^{x_P]}
\ket{\psi_n^{(\alpha_t)}}
\end{equation*}
or simply as
\begin{equation}
\ket{\psi_n^{(\alpha_t)} \,;\,
     n_{a,1} , n_{a,2} , n_{b,1} , n_{b,2} \,;\,
     \frac{P}{2}, k - \frac{P}{2}}
\label{basisstates}
\end{equation}
where $k = 0,1,\dots,P$.
For a fixed $N = n_{a,1} + n_{a,2} + n_{b,1} + n_{b,2}$, these states
transform in the $\left( \frac{N+P}{2} , \frac{N}{2} \right)$ representation
under the $SU(2)_T \times SU(2)_A$ subgroup. They are, in general, not eigenstates of $J$.
The $(P+1)$ states of the form
\begin{equation*}
\ket{\psi_0^{(P+\frac{3}{2})} \,;\,
     0 , 0 , 0 , 0 \,;\,
     \frac{P}{2}, k - \frac{P}{2}}
\qquad \qquad
(k = 0,1,\dots,P)
\end{equation*}
that transform in the $\left( \frac{P}{2} , 0 \right)$ representation of $SU(2)_T
\times SU(2)_A$ are, however, all eigenstates of $J$  with  eigenvalue $\mathfrak{J} = - \frac{P}{2}$. These $P+1$ states are annihilated by grade $-1$ operators
in $\mathfrak{C}^-_D$ of $\mathfrak{so}^*(8)_D$ (given in equation
(\ref{SO*(8)DGr-1})).
The action of the coset generators
\begin{equation*}
SU(4) \,/\, \left[ \, SU(2)_T \times SU(2)_A \times U(1)_J \, \right]
\end{equation*}
given in equation (\ref{SU(4)coset}) on the above states leads to a set of
states transforming in an irreducible representation of $SU(4)$ with Dynkin
labels $(2t,0,0) = (P,0,0)$.
We denote this set of states as  $\ket{\Omega^{(P+\frac{3}{2})}(P,0,0)}$.
They are all eigenstates of $H$ ($AdS_7$ energy), with the lowest  eigenvalue of $E = t + 2 = \frac{P}{2} + 2$, and are annihilated by all grade $-1$ operators.
Therefore they form a lowest energy K-type and  uniquely define a positive energy unitary irreducible
representation of $SO^*(8)$, labeled by the $SU(2)_G$ spin $g = t =
\frac{P}{2}$. The resulting unitary irreducible representations  correspond to  deformations of the minimal unitary representation. With respect to  the $SU(2)_T \times
SU(2)_A \times U(1)_J$ subgroup of $SU(4)$, the states $\ket{\Omega^{(P+\frac{3}{2})}(P,0,0)}$
has the following
decomposition:
\begin{equation}
\ket{\Omega^{(P+\frac{3}{2})}(P,0,0)}
= \left( \frac{P}{2} , 0 \right)^{-\frac{P}{2}} \oplus
  \left( \frac{P}{2} - \frac{1}{2}, \frac{1}{2} \right)^{-\frac{P}{2}+1} \oplus
  \left( \frac{P}{2} - 1 , 1 \right)^{-\frac{P}{2}+2} \oplus
  \dots \dots \oplus
  \left( 0 , \frac{P}{2} \right)^{+\frac{P}{2}}
\end{equation}
where we have labeled the irreps of
 $SU(2)_T \times SU(2)_A \times U(1)_J $ as $\left( t , a \right)^{\mathfrak{J}}$.
All the other states of the ``particle basis'' of the deformed minrep can be
obtained from $\ket{\Omega^{(P+\frac{3}{2})}(P,0,0)}$ by repeatedly acting on them with the
six  operators in $\mathfrak{C}^+_D$ subspace of $SO^*(8)_D$.
These six operators $Y^m$, $Z^m$, $N_+$ and $B_+$ in $\mathfrak{C}^+_D$
transform under $SU(2)_T \times SU(2)_A \times U(1)_J$ as follows:
\begin{equation*}
6 = ( 1/2 , 1/2 )^0 \oplus
    ( 0 , 0 )^{+1} \oplus
    ( 0 , 0 )^{-1} \,.
\end{equation*}
The generators $\left( Y^1 , Z^1 \right)$ and $\left( Y^2 , Z^2 \right)$ form
two doublets under $SU(2)_T$, and  the generators $\left( Y^1 , Y^2 \right)$
and $\left( Z^1 , Z^2 \right)$ form two doublets under $SU(2)_A$. $N_+$ and $B_+$ are both singlets under $SU(2)_T$ and $SU(2)_A$. The generators $Y^m$ and $Z^m$ have zero
$J$-charge, while the generators $N_+$ and $B_+$ have $J$-charges $+1$ and $-1$,
respectively.

In Table \ref{Table:SO*(8)Dminrep}, we give the $SU(4) \times U(1)_H$
decomposition of the deformed minreps of $SO^*(8)$ uniquely determined by the $(P+1)$ states
\begin{equation}
\ket{\psi_0^{(P+\frac{3}{2})} \,;\,
     0 , 0 , 0 , 0 \,;\,
     \frac{P}{2} , k - \frac{P}{2}}
\qquad \qquad
k = 0,1,\dots,P \,.
\end{equation}
and labeled by the $SU(2)_G$ spin $g = t = \frac{P}{2}$.

\begin{small}
\begin{longtable}[c]{|r||c|c|}
\kill

\caption[The deformed minimal unitary representation of $SO^*(8)$]
{In this table, we give the $SU(4) \times U(1)_H$ decomposition of the deformed
minimal unitary representation of $SO^*(8)$, defined by the ``lowest weight state''
$\ket{\psi_0^{(P+\frac{3}{2})} \,;\, 0 , 0 , 0 , 0 \,;\, \frac{P}{2} , -
\frac{P}{2}}$ for any non-negative $P$. These are massless representations of $SO^*(8)$, considered as the
$6D$ conformal group. As massless $6D$ conformal fields, their $SU^*(4)$
transformations coincide with $SU(4)$ transformations of the lowest energy
K-type $\ket{\Omega^{(P+\frac{3}{2})}}$ and  whose eigenvalue $E$ of $H$ is the negative of the conformal  dimension $\ell$. First column
gives the states, second column gives the energy eigenvalues, and third
column gives the $SU(4)$ Dynkin labels.
\label{Table:SO*(8)Dminrep}} \\
\hline
& & \\
States~~~~~~~~ & $E $ & $SU(4)$ \\
       &           & Dynkin \\
& & \\
\hline
& & \\
\endfirsthead
\caption[]{(continued)} \\
\hline
& & \\
State~~~~~~~~ & $E=-\ell$ & $SU(4)$ \\
       &           & Dynkin \\
& & \\
\hline
& & \\
\endhead
& & \\
\hline
\endfoot
& & \\
\hline
\endlastfoot

$\ket{\Omega^{(P+\frac{3}{2})}(P,0,0)}$
 & $\frac{P}{2}+2$ & $(P,0,0)$ \\[8pt]

\hline
& & \\

$\mathfrak{C}^+_D \ket{\Omega^{(P+\frac{3}{2})}(P,0,0)}$
 & $\frac{P}{2}+3$ & $(P,1,0)$
\\[8pt]

\hline
& & \\

$\left( \mathfrak{C}^+_D \right)^2 \ket{\Omega^{(P+\frac{3}{2})}(P,0,0)}$
 & $\frac{P}{2}+4$ & $(P,2,0)$
\\[8pt]

\hline
& & \\

\vdots~~~~~~~~ & \vdots & \vdots
\\[8pt]

\hline
& & \\

$\left( \mathfrak{C}^+_D \right)^n \ket{\Omega^{(P+\frac{3}{2})}(P,0,0)}$
 & $\frac{P}{2}+n+2$ & $(P,n,0)$
\\[8pt]

\hline
& & \\

\vdots~~~~~~~~ & \vdots & \vdots
\\[8pt]

\end{longtable}
\end{small}

The states  with $SU(4)$  Dynkin labels $\left( P , n , 0
\right)$ and $AdS$ energy $\frac{P}{2} + n + 2$ decompose
into the following   $SU(2)_T \times SU(2)_A \times
U(1)_J$ irreps labeled as  $\left( t , a
\right)^{\mathfrak{J}}$:
\begin{equation}
\begin{split}
\left( P , n , 0 \right)
&= \left( \frac{P}{2} , 0 \right)^{-\frac{P}{2}-n}_{P\ge0 , n\ge0}
\\
& \quad
   \oplus
   \left( \frac{P}{2} - \frac{1}{2} , \frac{1}{2} \right)^{-\frac{P}{2}-n+1}_{P\ge1 , n\ge0}
   \oplus
   \left( \frac{P}{2} + \frac{1}{2} , \frac{1}{2} \right)^{-\frac{P}{2}-n+1}_{P\ge0 , n\ge1}
\\
& \quad
   \oplus
   \left( \frac{P}{2} - 1 , 1 \right)^{-\frac{P}{2}-n+2}_{P\ge2 , n\ge0}
   \oplus
   \left( \frac{P}{2} , 0 \right)^{-\frac{P}{2}-n+2}_{P\ge1 , n\ge1}
   \oplus
   \left( \frac{P}{2} , 1 \right)^{-\frac{P}{2}-n+2}_{P\ge1 , n\ge1}
   \oplus
   \left( \frac{P}{2} + 1 , 1 \right)^{-\frac{P}{2}-n+2}_{P\ge0 , n\ge2}
\\
& \qquad \vdots
\\
& \quad
   \oplus
   \left( 0 , \frac{P}{2} \right)^{+\frac{P}{2}+n}_{P\ge0 , n\ge0}
\end{split}
\end{equation}
with the subscripts denoting  the allowed values of $P$ and $n$.

From the above table, it is clear
that the deformed minrep of $SO^*(8)$ with deformation parameter
$t$ is, in fact, the doubleton representation of $SO^*(8)$ whose lowest energy K-type
has the $SU(4)$  Young tableau
\begin{equation*}
\ket{\Omega} = | ~ \underbrace{\genrowbox}_{2 t = P} ~ \rangle
\end{equation*}
previously constructed by the oscillator method in
\cite{Gunaydin:1984wc,Gunaydin:1999ci}. We should stress again  that in the  oscillator construction of \cite{Gunaydin:1984wc,Gunaydin:1999ci}, one realizes the generators of $SO^*(8)$ as bilinears of two sets of twistorial bosonic oscillators transforming in the spinor representation of $SO^*(8)$, and the Fock space of these oscillators decomposes into the direct sum of infinitely many doubleton irreps of $SO^*(8)$. In the quasiconformal approach, each deformation labeled by $SU(2)_G$ spin ($g = t$) leads to a unique unitary irrep as explained above.


\section{Deformations of the  Minimal Unitary Representation  of $OSp(8^*|2N)$}
\label{OSp(8*|2N)D}

In our previous work \cite{Fernando:2010dp}, we constructed the ``undeformed'' minimal unitary
supermultiplet of $\mathfrak{osp}(8^*|2N)$,  with particular emphasis on that  of $\mathfrak{osp}(8^*|4)$. The minimal unitary supermultiplet of
$\mathfrak{osp}(8^*|4)$ is simply the supermultiplet of the six dimensional (2,0) conformal field
theory that is believed to be dual to the M-theory on $AdS_7 \times S^4$.

In this section, we will extend the deformed minimal unitary representations
of $SO^*(8)$, constructed in section \ref{SO*(8)D}, to deformed minimal unitary supermultiplets of $OSp(8^*|2N)$. For this purpose, in addition to the $\xi$- and $\chi$-type fermionic oscillators introduced for deforming the minrep of $SO^*(8)$,  we introduce a new set of fermionic oscillators required by supersymmetry. More specifically to realize the
compact Lie algebra $\mathfrak{usp}(2N)$, we introduce $N$ copies of $\alpha$- and
$\beta$-type fermionic oscillators (with indices $r = 1,2,\dots,N$), as reviewed in
Appendix \ref{USp(2N)} (see equation (\ref{susyfermions})). We shall refer to the fermions $\xi$ and $\chi$ as {\it deformation fermions}  and the fermions $\alpha$ and $\beta$ as {\it supersymmetry fermions}.

Recall that we labeled the $\mathfrak{su}(2)$ subalgebra that commutes with $\mathfrak{so}^*(4)$ subalgebra in the 5-graded decomposition  of  the minimal unitary realization of $\mathfrak{so}^*(8)$  as $\mathfrak{su}(2)_S$ such that  \[\mathfrak{g}^{(0)} =
\mathfrak{su}(2)_A \oplus \mathfrak{su}(1,1)_N \oplus \mathfrak{su}(2)_S
\oplus \mathfrak{so}(1,1) \,. \] Under deformation of the minimal unitary realization of $\mathfrak{so}^*(8)$,  the subalgebra
$\mathfrak{su}(2)_S$ gets contributions from deformation fermions $\xi$ and $\chi$  and goes over to $\mathfrak{su}(2)_T$, which is the diagonal
subalgebra of $\mathfrak{su}(2)_S$ and $\mathfrak{su}(2)_G$ (as defined in
equation (\ref{SU(2)G_gen})). Now  in extending the deformed minimal unitary realization of  $\mathfrak{so}^*(8)$ to the deformed minimal unitary realization of
$\mathfrak{osp}(8^*|2N)$,  it  gets further  contributions from supersymmetry fermions $\alpha$ and $\beta$.
 In particular  the subalgebra
$\mathfrak{su}(2)_T$ gets extended to $\mathfrak{su}(2)_{\mathcal{T}}$, which is the
diagonal subalgebra of $\mathfrak{su}(2)_T$ and $\mathfrak{su}(2)_F$, which involves only supersymmetry fermions $\alpha$ and $\beta$ (as
given in  equation (\ref{SU(2)F_gen})).
The generators of
$\mathfrak{su}(2)_{\mathcal{T}}$ turn out to be given by:
\begin{equation}
\begin{split}
\mathcal{T}_+
&= S_+ + F_+ + G_+
 = a^m b_m + \alpha^r \beta_r + \xi^x \chi_x
\\
\mathcal{T}_-
&= S_- + F_- + G_-
 = b^m a_m + \beta^r \alpha_r + \chi^x \xi_x
\\
\mathcal{T}_0
&= S_0 + F_0 + G_0
 = \frac{1}{2} \left( N_a - N_b + N_\alpha - N_\beta + N_\xi - N_\chi \right)
\end{split}
\end{equation}
The quadratic Casimir of  $\mathfrak{su}(2)_{\mathcal{T}}$ is
\begin{equation}
\mathcal{C}_2 \left[ \mathfrak{su}(2)_{\mathcal{T}} \right]
= \mathcal{T}^2
= \mathcal{T}_0 \mathcal{T}_0
  + \frac{1}{2}
    \left( \mathcal{T}_+ \mathcal{T}_- + \mathcal{T}_- \mathcal{T}_+ \right)
\,.
\end{equation}

Now  $\mathfrak{osp}(8^*|2N)_D$ has a 5-graded decomposition
\begin{equation}
\mathfrak{osp}(8^*|2N)_D
= \mathfrak{g}^{(-2)}_D \oplus
  \mathfrak{g}^{(-1)}_D \oplus
  \mathfrak{g}^{(0)}_D \oplus
  \mathfrak{g}^{(+1)}_D \oplus
  \mathfrak{g}^{(+2)}_D
\end{equation}
with respect to the subsuperalgebra
\begin{equation}
\mathfrak{g}^{(0)}_D
= \mathfrak{osp}(4^*|2N) \oplus
  \mathfrak{su}(2)_{\mathcal{T}} \oplus
  \mathfrak{so}(1,1)_\Delta
\end{equation}
such that grade $\pm2$ subspaces are one-dimensional.
In the deformed minimal unitary realization, generators of the subsuperalgebra $\mathfrak{osp}(4^*|2N)$ belonging to grade zero subspace do not get any contributions from the deformation fermions $\xi$ and $\chi$. Its generators have the same realization
as in the undeformed case. The generators of $\mathfrak{so}^*(4) =
\mathfrak{su}(2)_A \oplus \mathfrak{su}(1,1)_N$, those of
$\mathfrak{usp}(2N)$, and the $8N$ supersymmetry generators are given by:
\begin{equation}
\begin{aligned}
A_+ &= a^1 a_2 + b^1 b_2
\\
A_- &= \left( A_+ \right)^\dag = a_1 a^2 + b_1 b^2
\\
A_0 &= \frac{1}{2} \left( a^1 a_1 - a^2 a_2 + b^1 b_1 - b^2 b_2 \right)
\end{aligned}
\qquad  \qquad
\begin{aligned}
N_+ &= a^1 b^2 - a^2 b^1
\\
N_- &= \left( N_+ \right)^\dag = a_1 b_2 - a_2 b_1
\\
N_0 &= \frac{1}{2} \left( N_a + N_b \right) + 1
\end{aligned}
\end{equation}
\begin{equation}
\begin{split}
S_{rs}
&= \alpha_r \beta_s + \alpha_s \beta_r
\\
M^r_{~s}
&= \alpha^r \alpha_s - \beta_s \beta^r
\\
S^{rs}
&= \beta^r \alpha^s + \beta^s \alpha^r
 = \left( S_{rs} \right)^\dag
\end{split}
\end{equation}
\begin{equation}
\begin{aligned}
\Pi_{mr}
&= a_m \beta_r - b_m \alpha_r
\\
\Sigma_m^{~r}
&= a_m \alpha^r + b_m \beta^r
\end{aligned}
\qquad \qquad
\begin{aligned}
\overline{\Pi}^{mr}
&= \left( \Pi_{mr} \right)^\dag
 = a^m \beta^r - b^m \alpha^r
\\
\overline{\Sigma}^m_{~r}
&= \left( \Sigma_m^{~r} \right)^\dag
 = a^m \alpha_r + b^m \beta_r
\end{aligned}
\end{equation}

In the undeformed minimal unitary
realization, the quadratic Casimir of the subsuperalgebra
$\mathfrak{osp}(4^*|2N)$ turns out to be equal to the quadratic Casimir of the diagonal
$\mathfrak{su}(2)$ subalgebra of $\mathfrak{su}(2)_S$ and $\mathfrak{su}(2)_F$, that commutes with $\mathfrak{so}^*(4)$, modulo
an additive constant that depends on $N$ \cite{Fernando:2010dp}.
However,  in the deformed realization, no such simple relation holds between the
quadratic Casimir of $\mathfrak{osp}(4^*|2N)$ and that of
$\mathfrak{su}(2)_{\mathcal{T}}$.  In the deformed minimal
unitary realization, the quadratic Casimir of the subsuperalgebra
$\mathfrak{osp}(4^*|2N)$ is given by
\begin{equation}
\begin{split}
\mathcal{C}_2 \left[ \mathfrak{osp}(4^*|2N) \right]
&= \mathcal{C}_2 \left[ \mathfrak{su}(2)_{\mathcal{T}} \right]
   + \mathcal{C}_2 \left[ \mathfrak{su}(2)_G \right]
   - 2 \left[
        \mathcal{T}_0 G_0
        + \frac{1}{2} \left( \mathcal{T}_+ G_- + \mathcal{T}_- G_+ \right)
       \right]
\\
&\qquad
   - \frac{N \left( N - 4 \right)}{16} \,.
\end{split}
\end{equation}

The single bosonic generator in grade $-2$ subspace and  8 bosonic
operators plus $4N$ supersymmetry generators in grade $-1$ subspace of the
deformed minimal realization of $\mathfrak{osp}(8^*|2N)_D$ are also unchanged:
\begin{equation}
\mathcal{K}_- = K_- = \frac{1}{2} x^2
\end{equation}
\begin{equation}
\begin{aligned}
\mathcal{U}_m &= U_m = x \, a_m
\\
\mathcal{V}_m &= V_m = x \, b_m
\end{aligned}
\qquad \qquad \qquad \qquad
\begin{aligned}
\mathcal{U}^m &= U^m = x \, a^m
\\
\mathcal{V}^m &= V^m = x \, b^m
\end{aligned}
\end{equation}
\begin{equation}
\begin{aligned}
\mathcal{Q}_r &= Q_r = x \, \alpha_r
\\
\mathcal{S}_r &= S_m = x \, \beta_r
\end{aligned}
\qquad \qquad \qquad \qquad
\begin{aligned}
\mathcal{Q}^r &= Q^r = x \, \alpha^m
\\
\mathcal{S}^r &= S^r = x \, \beta^m
\end{aligned}
\end{equation}
They form a super Heisenberg algebra by (anti-)commuting into
$\mathcal{K}_-$.
However, since $\mathfrak{su}(2)_S$ has now been extended to
$\mathfrak{su}(2)_{\mathcal{T}}$,
the grade $+2$ generator now depends on $\mathcal{T}^2$ and is given by
\begin{equation}
\mathcal{K}_+
= \frac{1}{2} p^2
  + \frac{1}{4 \, x^2} \left( 8 \, \mathcal{T}^2 + \frac{3}{2} \right) \,.
\end{equation}
Therefore the generators in grade $+1$ subspace get modified, since they are obtained from the commutators of the form
$\commute{\mathfrak{g}_D^{(-1)}}{\mathfrak{g}_D^{(+2)}}$:
\begin{equation}
\begin{split}
\widetilde{\mathcal{U}}_m
= i \commute{\,\mathcal{U}_m}{\mathcal{K}_+}
& \qquad \qquad \qquad \qquad
\widetilde{\mathcal{U}}^m
= \left( \widetilde{\mathcal{U}}_m \right)^\dag
= i \commute{\,\mathcal{U}^m}{\mathcal{K}_+}
\\
\widetilde{\mathcal{V}}_m
= i \commute{\,\mathcal{V}_m}{\mathcal{K}_+}
& \qquad \qquad \qquad \qquad
\widetilde{\mathcal{V}}^m
= \left( \widetilde{\mathcal{V}}_m \right)^\dag
= i \commute{\,\mathcal{V}^m}{\mathcal{K}_+}
\end{split}
\end{equation}
\begin{equation}
\begin{split}
\widetilde{\mathcal{Q}}_r
= i \commute{\,\mathcal{Q}_r}{\mathcal{K}_+}
& \qquad \qquad \qquad \qquad
\widetilde{\mathcal{Q}}^r
= \left( \widetilde{\mathcal{Q}}_r \right)^\dag
= i \commute{\,\mathcal{Q}^r}{\mathcal{K}_+}
\\
\widetilde{\mathcal{S}}_r
= i \commute{\,\mathcal{S}_r}{\mathcal{K}_+}
& \qquad \qquad \qquad \qquad
\widetilde{\mathcal{S}}^r
= \left( \widetilde{\mathcal{S}}_r \right)^\dag
= i \commute{\,\mathcal{S}^r}{\mathcal{K}_+}
\end{split}
\end{equation}
The explicit form of these 8 bosonic generators and $4N$ supersymmetry
generators of grade $+1$ subspace are as follows:
\begin{equation}
\begin{split}
\widetilde{\mathcal{U}}_m
&= - p \, a_m
   + \frac{2i}{x}
     \left[
      \left( \mathcal{T}_0 + \frac{3}{4} \right) a_m + \mathcal{T}_- b_m
     \right]
\\
\widetilde{\mathcal{U}}^m
&= - p \, a^m
   - \frac{2i}{x}
     \left[
      \left( \mathcal{T}_0 - \frac{3}{4} \right) a^m + \mathcal{T}_+ b^m
     \right]
\\
\widetilde{\mathcal{V}}_m
&= - p \, b_m
   - \frac{2i}{x}
     \left[
      \left( \mathcal{T}_0 - \frac{3}{4} \right) b_m - \mathcal{T}_+ a_m
     \right]
\\
\widetilde{\mathcal{V}}^m
&= - p \, b^m
   + \frac{2i}{x}
     \left[
      \left( \mathcal{T}_0 + \frac{3}{4} \right) b^m - \mathcal{T}_- a^m
     \right]
\end{split}
\end{equation}
\begin{equation}
\begin{split}
\widetilde{\mathcal{Q}}_r
&= - p \, \alpha_r
   + \frac{2i}{x}
     \left[
      \left( \mathcal{T}_0 + \frac{3}{4} \right) \alpha_r
      + \mathcal{T}_- \beta_r
     \right]
\\
\widetilde{\mathcal{Q}}^r
&= - p \, \alpha^r
   - \frac{2i}{x}
     \left[
      \left( \mathcal{T}_0 - \frac{3}{4} \right) \alpha^r
      + \mathcal{T}_+ \beta^r
     \right]
\\
\widetilde{\mathcal{S}}_r
&= - p \, \beta_r
   - \frac{2i}{x}
     \left[
      \left( \mathcal{T}_0 - \frac{3}{4} \right) \beta_r
      - \mathcal{T}_+ \alpha_r
     \right]
\\
\widetilde{\mathcal{S}}^r
&= - p \, \beta^r
   + \frac{2i}{x}
     \left[
      \left( \mathcal{T}_0 + \frac{3}{4} \right) \beta^r
      - \mathcal{T}_- \alpha^r
     \right]
\end{split}
\end{equation}
These grade $+1$ generators (anti-)commute into the grade $+2$ generator and
form a super Heisenberg algebra.

The anticommutators between the supersymmetry generators in
$\mathfrak{g}^{(-1)}_D$ and  $\mathfrak{g}^{(+1)}_D$ given above close into
the bosonic generators in $\mathfrak{g}^{(0)}_D$:
\begin{equation}
\begin{aligned}
\anticommute{\mathcal{Q}_r}{\widetilde{\mathcal{Q}}_s}
&= 0
\\
\anticommute{\mathcal{Q}_r}{\widetilde{\mathcal{Q}}^s}
&= - \delta^s_r \, \Delta - 2 i \, \delta^s_r \, \mathcal{T}_0
   + 2 i \, M^s_{~r}
\\
\anticommute{\mathcal{S}_r}{\widetilde{\mathcal{S}}_s}
&= 0
\\
\anticommute{\mathcal{S}_r}{\widetilde{\mathcal{S}}^s}
&= - \delta^s_r \, \Delta + 2 i \, \delta^s_r \, \mathcal{T}_0
   + 2 i \, M^s_{~r}
\end{aligned}
\qquad \qquad
\begin{aligned}
\anticommute{\mathcal{Q}_r}{\widetilde{\mathcal{S}}_s}
&= - 2 i \, S_{rs}
\\
\anticommute{\mathcal{Q}_r}{\widetilde{\mathcal{S}}^s}
&= - 2 i \, \delta^s_r \, \mathcal{T}_-
\\
\anticommute{\mathcal{S}_r}{\widetilde{\mathcal{Q}}_s}
&= + 2 i \, S_{rs}
\\
\anticommute{\mathcal{S}_r}{\widetilde{\mathcal{Q}}^s}
&= - 2 i \, \delta^s_r \, \mathcal{T}_+
\end{aligned}
\end{equation}

The commutators between the bosonic (even) and fermionic (odd) generators of
$\mathfrak{g}^{(-1)}_D$ and $\mathfrak{g}^{(+1)}_D$ subspaces close into the
fermionic (odd) generators of $\mathfrak{g}^{(0)}_D$:
\begin{equation}
\begin{aligned}
\commute{\mathcal{U}_m}{\widetilde{\mathcal{Q}}_r}
&= 0
\\
\commute{\mathcal{U}_m}{\widetilde{\mathcal{Q}}^r}
&= - 2 i \, \Sigma_m^{~r}
\\
\commute{\mathcal{U}_m}{\widetilde{\mathcal{S}}_r}
&= - 2 i \, \Pi_{mr}
\\
\commute{\mathcal{U}_m}{\widetilde{\mathcal{S}}^r}
&= 0
\end{aligned}
\qquad \qquad
\begin{aligned}
\commute{\mathcal{V}_m}{\widetilde{\mathcal{Q}}_r}
&= + 2 i \, \Pi_{mr}
\\
\commute{\mathcal{V}_m}{\widetilde{\mathcal{Q}}^r}
&= 0
\\
\commute{\mathcal{V}_m}{\widetilde{\mathcal{S}}_r}
&= 0
\\
\commute{\mathcal{V}_m}{\widetilde{\mathcal{S}}^r}
&= - 2 i \, \Sigma_m^{~r}
\end{aligned}
\end{equation}
\begin{equation}
\begin{aligned}
\commute{\mathcal{Q}_r}{\widetilde{\mathcal{U}}_m}
&= 0
\\
\commute{\mathcal{Q}^r}{\widetilde{\mathcal{U}}_m}
&= - 2 i \, \Sigma_m^{~r}
\\
\commute{\mathcal{S}_r}{\widetilde{\mathcal{U}}_m}
&= - 2 i \, \Pi_{mr}
\\
\commute{\mathcal{S}^r}{\widetilde{\mathcal{U}}_m}
&= 0
\end{aligned}
\qquad \qquad
\begin{aligned}
\commute{\mathcal{Q}_r}{\widetilde{\mathcal{V}}_m}
&= + 2 i \, \Pi_{mr}
\\
\commute{\mathcal{Q}^r}{\widetilde{\mathcal{V}}_m}
&= 0
\\
\commute{\mathcal{S}_r}{\widetilde{\mathcal{V}}_m}
&= 0
\\
\commute{\mathcal{S}^r}{\widetilde{\mathcal{V}}_m}
&= - 2 i \, \Sigma_m^{~r}
\end{aligned}
\end{equation}

Thus the 5-grading of the Lie superalgebra $\mathfrak{osp}(8^*|2N)_D$,
defined by the  generator $\Delta$, takes the form:
\begin{equation}
\begin{split}
\mathfrak{osp}(8^*|4)_D
&= \mathfrak{g}^{(-2)}_D \oplus
   \mathfrak{g}^{(-1)}_D \oplus
   \left[ \mathfrak{osp}(4^*|2N) \oplus
          \mathfrak{su}(2)_{\mathcal{T}} \oplus
          \mathfrak{so}(1,1)_\Delta
   \right] \oplus
   \mathfrak{g}^{(+1)}_D \oplus
   \mathfrak{g}^{(+2)}_D
\\
&= \mathcal{K}_-
   \oplus
   \left[ \mathcal{U}_m \,,\, \mathcal{U}^m \,,\,
          \mathcal{V}_m \,,\, \mathcal{V}^m \,,\,
          \mathcal{Q}_r \,,\, \mathcal{Q}^r \,,\,
          \mathcal{S}_r \,,\, \mathcal{S}^r \right]
\\
& \qquad
   \oplus
   \left[ A_{\pm,0} \,,\, N_{\pm,0} \,,\,
          S_{rs} \,,\, M^r_{~s} \,,\, S^{rs} \,,\,
          \Pi_{mr} \,,\, \overline{\Pi}^{mr} \,,\,
          \Sigma_m^{~r} \,,\, \overline{\Sigma}^m_{~r} \,,\,
          \mathcal{T}_{\pm,0} \,,\, \Delta
   \right]
\\
& \qquad \qquad
   \oplus
   \left[ \widetilde{\mathcal{U}}_m \,,\,
          \widetilde{\mathcal{U}}^m \,,\,
          \widetilde{\mathcal{V}}_m \,,\,
          \widetilde{\mathcal{V}}^m \,,\,
          \widetilde{\mathcal{Q}}_r \,,\,
          \widetilde{\mathcal{Q}}^r \,,\,
          \widetilde{\mathcal{S}}_r \,,\,
          \widetilde{\mathcal{S}}^r
   \right]
   \oplus
   \mathcal{K}_+
\end{split}
\end{equation}

The quadratic Casimir of $\mathfrak{osp}(8^*|2N)_D$ is given by
\begin{equation}
\begin{split}
\mathcal{C}_2 \left[ \mathfrak{osp}(8^*|2N)_D \right]
&= \mathcal{C}_2 \left[ \mathfrak{osp}(4^*|2N) \right]
   - \mathcal{C}_2 \left[ \mathfrak{su}(2)_\mathcal{T} \right]
   + \mathcal{C}_2 \left[ \mathfrak{su}(1,1)_\mathcal{K} \right]
\\
& \quad
   - \frac{i}{8} \mathcal{F}_1 \left( \mathcal{U} , \mathcal{V} \right)
   + \frac{i}{8} \mathcal{F}_2 \left( \mathcal{Q} , \mathcal{S} \right)
\end{split}
\end{equation}
where
\begin{equation}
\begin{split}
\mathcal{F}_1 \left( \mathcal{U} , \mathcal{V} \right)
&= \left( \mathcal{U}_m \widetilde{\mathcal{U}}^m
          + \mathcal{V}_m \widetilde{\mathcal{V}}^m
          + \widetilde{\mathcal{U}}^m \mathcal{U}_m
          + \widetilde{\mathcal{V}}^m \mathcal{V}_m \right)
\\
& \qquad
   - \left( \mathcal{U}^m \widetilde{\mathcal{U}}_m
            + \mathcal{V}^m \widetilde{\mathcal{V}}_m
            + \widetilde{\mathcal{U}}_m \mathcal{U}^m
            + \widetilde{\mathcal{V}}_m \mathcal{V}^m \right)
\\
\mathcal{F}_2 \left( \mathcal{Q} , \mathcal{S} \right)
&= \left( \mathcal{Q}_r \widetilde{\mathcal{Q}}^r
          + \mathcal{S}_r \widetilde{\mathcal{S}}^r
          - \widetilde{\mathcal{Q}}^r \mathcal{Q}_r
          - \widetilde{\mathcal{S}}^r \mathcal{S}_r \right)
\\
& \qquad
   + \left( \mathcal{Q}^r \widetilde{\mathcal{Q}}_r
            + \mathcal{S}^r \widetilde{\mathcal{S}}_r
            - \widetilde{\mathcal{Q}}_r \mathcal{Q}^r
            - \widetilde{\mathcal{S}}_r \mathcal{S}^r \right)
\end{split}
\end{equation}
and reduces to
\begin{equation}
\mathcal{C}_2 \left[ \mathfrak{osp}(8^*|2N)_D \right]
= G^2 - \frac{1}{16} \left( N^2 - 20 \, N + 32 \right)
\label{QC_OSp(8*|2N)D}
\end{equation}
where $G^2$ is the quadratic Casimir of $\mathfrak{su}(2)_G$. Hence each deformed irreducible minimal
unitary supermultiplet of $\mathfrak{osp}(8^*|2N)$ can be  labeled by the eigenvalues $g(g+1)$ of $G^2$ as we shall explicitly show later.


\section{The Compact 3-Grading of $OSp(8^*|2N)_D$}
\label{3GrOSp(8*|2N)D}

The Lie superalgebra $\mathfrak{osp}(8^*|2N)_D$ can be given a 3-graded
decomposition with respect to its compact subsuperalgebra $\mathfrak{u}(4|N)
= \mathfrak{su}(4|N) \oplus \mathfrak{u}(1)$:
\begin{equation}
\mathfrak{osp}(8^*|2N)_D
= \mathfrak{C}^-_D \oplus \mathfrak{C}^0_D \oplus \mathfrak{C}^+_D
\end{equation}
where
\begin{equation}
\begin{split}
\mathfrak{C}^-_D
&= \frac{1}{2} \left( \mathcal{U}_m - i \, \widetilde{\mathcal{U}}_m \right)
   \oplus
   \frac{1}{2} \left( \mathcal{V}_m - i \, \widetilde{\mathcal{V}}_m \right)
   \oplus
   N_-
   \oplus
   \frac{i}{2}
    \left[ \Delta + i \left( \mathcal{K}_+ - \mathcal{K}_- \right) \right]
   \oplus
   S_{rs}
\\
& \quad \oplus
   \frac{1}{2} \left( \mathcal{Q}_r - i \, \widetilde{\mathcal{Q}}_r \right)
   \oplus
   \frac{1}{2} \left( \mathcal{S}_r - i \, \widetilde{\mathcal{S}}_r \right)
   \oplus
   \Pi_{mr}
\\
\mathfrak{C}^0_D
&= \left[
    \mathcal{T}_{\pm,0} \oplus
    A_{\pm,0} \oplus
    \left[
     N_0 - \frac{1}{2} \left( \mathcal{K}_+ + \mathcal{K}_- \right)
    \right]
    \oplus
    \frac{1}{2} \left( \mathcal{U}_m + i \, \widetilde{\mathcal{U}}_m \right)
    \oplus
    \frac{1}{2} \left( \mathcal{U}^m - i \, \widetilde{\mathcal{U}}^m \right)
   \right.
\\
& \qquad
   \left. \oplus
    \frac{1}{2} \left( \mathcal{V}_m + i \, \widetilde{\mathcal{V}}_m \right)
    \oplus
    \frac{1}{2} \left( \mathcal{V}^m - i \, \widetilde{\mathcal{V}}^m \right)
    \oplus
    M^r_{~s} \oplus
    \left[
     \frac{1}{2} \left( \mathcal{K}_+ + \mathcal{K}_- \right)
     + \frac{2}{N} M_0
    \right]
   \right]
   \oplus
   \mathfrak{H}
\\
& \quad \oplus
   \frac{1}{2} \left( \mathcal{Q}_r + i \, \widetilde{\mathcal{Q}}_r \right)
   \oplus
   \frac{1}{2} \left( \mathcal{Q}^r - i \, \widetilde{\mathcal{Q}}^r \right)
   \oplus
   \frac{1}{2} \left( \mathcal{S}_r + i \, \widetilde{\mathcal{S}}_r \right)
   \oplus
   \frac{1}{2} \left( \mathcal{S}^r - i \, \widetilde{\mathcal{S}}^r \right)
   \oplus
   \Sigma_m^{~r} \oplus
   \overline{\Sigma}^m_{~r}
\\
\mathfrak{C}^+_D
&= \frac{1}{2} \left( \mathcal{U}^m + i \, \widetilde{\mathcal{U}}^m \right)
   \oplus
   \frac{1}{2} \left( \mathcal{V}^m + i \, \widetilde{\mathcal{V}}^m \right)
   \oplus
   N_+
   \oplus
   - \frac{i}{2}
     \left[ \Delta - i \left( \mathcal{K}_+ - \mathcal{K}_- \right) \right]
   \oplus
   S^{rs}
\\
& \quad \oplus
   \frac{1}{2} \left( \mathcal{Q}^r + i \, \widetilde{\mathcal{Q}}^r \right)
   \oplus
   \frac{1}{2} \left( \mathcal{S}^r + i \, \widetilde{\mathcal{S}}^r \right)
   \oplus
   \overline{\Pi}^{mr}
\end{split}
\label{OSp(8*|2N)3Gr}
\end{equation}

The $\mathfrak{u}(1)$ generator $\mathfrak{H}$ that defines the compact 3-grading of
$\mathfrak{osp}(8^*|2N)_D$ is given by
\begin{equation}
\begin{split}
\mathfrak{H}
&= \frac{1}{2} \left( \mathcal{K}_+ + \mathcal{K}_- \right)
    + N_0 + M_0
\\
&= \frac{1}{4} \left( x^2 + p^2 \right)
    + \frac{1}{x^2}
      \left( \mathcal{T}^2 + \frac{3}{16} \right)
    + \frac{1}{2} \left( N_a + N_b + N_\alpha + N_\beta \right)
    + \frac{2 - N}{2}
\end{split}
\label{deformed3-GrGenerator}
\end{equation}
and  plays the role of the ``total energy'' operator for  the deformed
minrep of $\mathfrak{osp}(8^*|2N)$.

In the supersymmetric extension of the deformed minrep, the $\mathfrak{u}(1)$
generator that corresponds to the $AdS_7$ energy and determines a 3-grading
of $\mathfrak{so}^*(8)_D$ is given by
\begin{equation}
\begin{split}
\mathcal{H}
&= \frac{1}{2} \left( \mathcal{K}_+ + \mathcal{K}_- \right) + N_0
\\
&= \frac{1}{4} \left( x^2 + p^2 \right)
   + \frac{1}{x^2}
      \left( \mathcal{T}^2 + \frac{3}{16} \right)
   + \frac{1}{2} \left( N_a + N_b \right) + 1
\\
&= \mathcal{H}_\odot + H_a + H_b
\end{split}
\label{deformedNonSusyH}
\end{equation}
where $\mathcal{H}_\odot$ is the Hamiltonian of the singular oscillator:
\begin{equation}
\mathcal{H}_\odot
= \frac{1}{2} \left( \mathcal{K}_+ + \mathcal{K}_- \right)
= \frac{1}{4} \left( x^2 + p^2 \right)
  + \frac{1}{x^2}
     \left( \mathcal{T}^2 + \frac{3}{16} \right)
\end{equation}
and $H_a$ and $H_b$ are the Hamiltonians corresponding to $a$- and $b$-type
bosonic oscillators, respectively:
\begin{equation*}
H_a = \frac{1}{2} \left( N_a + 1 \right)
\qquad \qquad \qquad
H_b = \frac{1}{2} \left( N_b + 1 \right)
\end{equation*}

We shall label the bosonic operators that belong to the
subspace $\mathfrak{C}^-_D$ of the deformed $\mathfrak{osp}(8^*|2N)_D$ in the
compact 3-grading  as follows:
\begin{equation}
\begin{split}
\mathcal{Y}_m
&= \frac{1}{2} \left( \mathcal{U}_m - i \, \widetilde{\mathcal{U}}_m \right)
 = \frac{1}{2} \left( x + i \, p \right) a_m
   + \frac{1}{x}
     \left[
      \left( \mathcal{T}_0 + \frac{3}{4} \right) a_m + \mathcal{T}_- b_m
     \right]
\\
\mathcal{Z}_m
&= \frac{1}{2} \left( \mathcal{V}_m - i \, \widetilde{\mathcal{V}}_m \right)
 = \frac{1}{2} \left( x + i \, p \right) b_m
   - \frac{1}{x}
     \left[
      \left( \mathcal{T}_0 - \frac{3}{4} \right) b_m - \mathcal{T}_+ a_m
     \right]
\\
N_-
&= a_1 b_2 - a_2 b_1
\\
\mathcal{B}_-
&= \frac{i}{2}
    \left[ \Delta + i \left( \mathcal{K}_+ - \mathcal{K}_- \right) \right]
 = \frac{1}{4} \left( x + i \, p \right)^2
   - \frac{1}{x^2}
      \left( \mathcal{T}^2 + \frac{3}{16} \right)
\\
S_{rs}
&= \alpha_r \beta_s + \alpha_s \beta_r
\end{split}
\label{deformedOSp(8*|N)Gr-1B}
\end{equation}
and the  $4N$ supersymmetry generators in $\mathfrak{C}^-_D$ subspace as:
\begin{equation}
\begin{split}
\mathfrak{Q}_r
&= \frac{1}{2} \left( \mathcal{Q}_r - i \, \widetilde{\mathcal{Q}}_r \right)
 = \frac{1}{2} \left( x + i \, p \right) \alpha_r
   + \frac{1}{x}
     \left[
      \left( \mathcal{T}_0 + \frac{3}{4} \right) \alpha_r
      + \mathcal{T}_- \beta_r
     \right]
\\
\mathfrak{S}_r
&= \frac{1}{2} \left( \mathcal{S}_r - i \, \widetilde{\mathcal{S}}_r \right)
 = \frac{1}{2} \left( x + i \, p \right) \beta_r
   - \frac{1}{x}
     \left[
      \left( \mathcal{T}_0 - \frac{3}{4} \right) \beta_r
      - \mathcal{T}_+ \alpha_r
     \right]
\\
\Pi_{mr}
&= a_m \beta_r - b_m \alpha_r
\end{split}
\label{deformedOSp(8*|N)Gr-1F}
\end{equation}
The generators that belong to $\mathfrak{C}^+_D$ subspace are the Hermitian
conjugates of those in $\mathfrak{C}^-_D$. Then the bosonic operators in
$\mathfrak{C}^+_D$ are:
\begin{equation}
\begin{split}
\mathcal{Y}^m
&= \frac{1}{2} \left( \mathcal{U}^m + i \, \widetilde{\mathcal{U}}^m \right)
 = \frac{1}{2} \left( x - i \, p \right) a^m
   + \frac{1}{x}
     \left[
      \left( \mathcal{T}_0 - \frac{3}{4} \right) a^m + \mathcal{T}_+ b^m
     \right]
\\
\mathcal{Z}^m
&= \frac{1}{2} \left( \mathcal{V}^m + i \, \widetilde{\mathcal{V}}^m \right)
 = \frac{1}{2} \left( x - i \, p \right) b^m
   - \frac{1}{x}
     \left[
      \left( \mathcal{T}_0 + \frac{3}{4} \right) b^m - \mathcal{T}_- a^m
     \right]
\\
N_+
&= a^1 b^2 - a^2 b^1
\\
\mathcal{B}_+
&= - \frac{i}{2}
     \left[ \Delta - i \left( \mathcal{K}_+ - \mathcal{K}_- \right) \right]
 = \frac{1}{4} \left( x - i \, p \right)^2
   - \frac{1}{x^2}
      \left( \mathcal{T}^2 + \frac{3}{16} \right)
\\
S^{rs}
&= \alpha^r \beta^s + \alpha^s \beta^r
\end{split}
\label{deformedOSp(8*|N)Gr+1B}
\end{equation}
and the $4 N$ supersymmetry generators in $\mathfrak{C}^+_D$ subspace are:
\begin{equation}
\begin{split}
\mathfrak{Q}^r
&= \frac{1}{2} \left( \mathcal{Q}^r + i \, \widetilde{\mathcal{Q}}^r \right)
 = \frac{1}{2} \left( x - i \, p \right) \alpha^r
   + \frac{1}{x}
     \left[
      \left( \mathcal{T}_0 - \frac{3}{4} \right) \alpha^r
      + \mathcal{T}_+ \beta^r
     \right]
\\
\mathfrak{S}^r
&= \frac{1}{2} \left( \mathcal{S}^r + i \, \widetilde{\mathcal{S}}^r \right)
 = \frac{1}{2} \left( x - i \, p \right) \beta^r
   - \frac{1}{x}
     \left[
      \left( \mathcal{T}_0 + \frac{3}{4} \right) \beta^r
      - \mathcal{T}_- \alpha^r
     \right]
\\
\overline{\Pi}^{mr}
&= a^m \beta^r - b^m \alpha^r
\end{split}
\label{deformedOSp(8*|N)Gr+1F}
\end{equation}

Once again, we have the important relation
\begin{equation}
\mathcal{Y}^1 \, \mathcal{Z}^2 - \mathcal{Y}^2 \, \mathcal{Z}^1
= N_+ \, \mathcal{B}_+
\end{equation}


\section{Deformed  Minimal Unitary Representations  of $OSp(8^*|2N)$ as $6D$ Massless Conformal Supermultiplets}
\label{deformedminrepsupermultiplets}

Since the quadratic Casimir of $OSp(8^*|2N)_D$ depends only on the quadratic Casimir of $SU(2)_G$ constructed out of deformation fermions $\xi$ and $\chi$ (see equation (\ref{QC_OSp(8*|2N)D})), just as the quadratic Casimir of $SO^*(8)_D$ depends only on the quadratic Casimir of $SU(2)_G$ (see equation (\ref{QC_SO*(8)D})), one expects to obtain an irreducible unitary supermultiplet of $OSp(8^*|2N)$ for each spin $g$ labeling the irreps of $SU(2)_G$.
Let us show that this indeed is the case.
For each $SU(2)_\mathcal{T}$ spin $\mathfrak{t} = g = \frac{P}{2} \ne 0$  there is a multiplet of states that are
annihilated by all the operators in grade $-1$ subspace
$\mathfrak{C}^-_D$ and transforms irreducibly under the subsuperalgebra $\mathfrak{u}(4|N)$,
which is the grade zero subspace $ \mathfrak{C}^0_D$. Let us call  this supermultiplet of states the ``lowest
energy K-type''  of $OSp(8^*|2N)_D$.
To obtain this lowest energy K-type for a given $\mathfrak{t}=g$,
consider the tensor product of states  of the form \[
\ket{\psi_n^{(\alpha_\mathfrak{t})} \,;\,
     n_{a,1} , n_{a,2} , n_{b,1} , n_{b,2} \,;\,
     \frac{P}{2}, k - \frac{P}{2}}
\] constructed earlier, with the states created by the supersymmetry fermions
\[ \alpha^{[r_1} \dots \alpha^{r_{n_\alpha}} \beta^{r_{n_\alpha+1}} \dots \beta^{r_{n_\alpha+n_\beta}]} \ket{0}_F
\]
where $\ket{0}_F$ is the fermionic Fock vacuum annihilated by all the fermionic annihilation operators $\xi_x$, $\chi_x$, $\alpha_r$ and $\beta_r$ ($x=1,\dots,P$ and $r=1,\dots,N$), and denote them as
\begin{equation}
\ket{\psi_n^{(\alpha_\mathfrak{t})} \,;\,
     n_{a,1} , n_{a,2} , n_{b,1} , n_{b,2} \,;\,
     n_\alpha , n_\beta \,;\,
     \frac{P}{2}, k - \frac{P}{2}} \,. \label{susybasisstates}
\end{equation}

The following $(P+1)$ states of the form
\begin{equation}
\ket{\psi_0^{(\alpha_\mathfrak{t})} \,;\,
     0 , 0 , 0 , 0 \,;\,
     0 , 0 \,;\,
     \frac{P}{2}, k - \frac{P}{2}}
\qquad \qquad
(k = 0,1,\dots,P)
\label{susyground}
\end{equation}
transform in the $\left( \frac{P}{2} , 0 \right)$ representation of
$SU(2)_\mathcal{T} \times SU(2)_A$ with a definite eigenvalue
$\mathfrak{J} = - \frac{P}{2}$ with respect to the $U(1)_\mathcal{J}$ generator $\mathcal{J} = N_0 - \frac{1}{2} \left( \mathcal{K}_+ + \mathcal{K}_- \right)$, and  are annihilated by the
six bosonic operators and $N \left( N + 1 \right) / 2$ supersymmetry
generators in grade $-1$ subspace $\mathfrak{C}^-_D$ of
$\mathfrak{osp}(8^*|2N)_D$ (given in equations
(\ref{deformedOSp(8*|N)Gr-1B}) and (\ref{deformedOSp(8*|N)Gr-1F})) for the choice
\begin{equation}
\alpha_\mathfrak{t} = 2 \mathfrak{t} + \frac{3}{2} = P +\frac{3}{2} \,.
\end{equation}
These states uniquely define a positive energy unitary supermultiplet of
$OSp(8^*|2N)$, labeled by the $SU(2)_G$ spin $g=\mathfrak{t} = \frac{P}{2}$,
which corresponds to a deformation of the minimal unitary supermultiplet.
By acting on the states in equation (\ref{susyground}) with the coset generators
\begin{equation*}
SU(4|N) \,/\, \left[ SU(2)_\mathcal{T} \times SU(2)_A \times U(1)_\mathcal{J} \right]
\end{equation*}
 one  obtains a set of states transforming in an
irreducible representation of $SU(4|N)$ with the Young supertableau $ \,\, \underbrace{\sgenrowbox}_{2 \mathfrak{t}} $.  We denote these states as
\[ \ket{\Omega^{(P+\frac{3}{2})} \,,\, \underbrace{\sgenrowbox}_{P=2\mathfrak{t}}} \]
For a given value of the deformation parameter $\mathfrak{t}$ ($= g$), they have the lowest ``total
energy'' $\mathfrak{H} = \mathfrak{t} + 1 = g + 1 = \frac{P}{2} + 1$ and are all annihilated by grade $-1$ operators.
Hence they form a lowest energy K-type of $OSp(8^*|2N)$.
By repeatedly acting on this set of states $\ket{\Omega^{(P+\frac{3}{2})} \,,\, \underbrace{\sgenrowbox}_{P=2\mathfrak{t}}}$ with the
operators in grade $+1$ subspace $\mathfrak{C}^+_D$, one obtains an infinite
set of states that form a basis of  a unitary irreducible
representation of $OSp(8^*|2N)$. This infinite set of states can be
decomposed into a finite number of irreducible representations of the even
subgroup $SO^*(8) \times USp(2N)$, with each irrep of $SO^*(8)$ corresponding
to a massless conformal field in six dimensions.

In Table \ref{Table:gendeformedsupermulN}, we present the general deformed minimal unitary supermultiplet of $\mathfrak{osp}(8^*|2N)$ corresponding to the deformation parameter $\mathfrak{t}$ ($= g$), obtained by starting from the lowest energy K-type
\[ \ket{\Omega^{(P+\frac{3}{2})} \,,\, \underbrace{\sgenrowbox}_{P=2\mathfrak{t}=2g}} \,. \]

\begin{small}
\begin{longtable}[c]{|c|c||l||c|}
\kill

\caption[The general deformed minimal unitary supermultiplet of
$\mathfrak{osp}(8^*|2N)$ corresponding to the deformation parameter $\mathfrak{t}$]
{The general deformed minimal unitary supermultiplet of
$\mathfrak{osp}(8^*|2N)$ corresponding to the deformation parameter $\mathfrak{t} = g = P/2$. First column  gives the $AdS$ energy $\mathcal{H}$ which is equal to negative conformal dimension $\ell$  of the corresponding conformal field. Last column gives the $SU(4)$ Dynkin labels of the lowest energy K-type of $SO^*(8)$, which coincides with the Dynkin labels of the corresponding conformal field under the Lorentz group $SU^*(4)$, and the Dynkin labels with respect to $USp(2N)$.
The decomposition of $SU(4)$ irreps with respect to $SU(2)_\mathcal{T} \times SU(2)_A
\times U(1)_\mathcal{J}$ is denoted by $\left( \mathfrak{t} , \mathfrak{a}
\right)^{\mathfrak{J}}$ and is listed in the third column. Second column gives the
eigenvalues of ``total energy'' $ \mathfrak{H}$.  Note that for $P<N$ the states with negative entries in their Dynkin labels do not occur.
\label{Table:gendeformedsupermulN}} \\
\hline
& & & \\
$\mathcal{H} = - \ell$ & $\mathfrak{H}$ &
$( \mathfrak{t} , \mathfrak{a} )^{\mathfrak{J}}$ & $SU^*(4)_{\mathrm{Dynkin}}=SU(4)_{\mathrm{Dynkin}} $ \\
&  &
 & $USp(2N)_{\mathrm{Dynkin}}$ \\
& & & \\
\hline
& & & \\
\endfirsthead
\caption[]{(continued)} \\
\hline
& & & \\
$\mathcal{H} = - \ell$ & $\mathfrak{H}$ &
$( \mathfrak{t} , \mathfrak{a} )^{\mathfrak{J}}$ & $SU^*(4)_{\mathrm{Dynkin}}=SU(4)_{\mathrm{Dynkin}} $ \\
&  &
 & $USp(2N)_{\mathrm{Dynkin}}$ \\
& & & \\
\hline
& & & \\
\endhead
& & & \\
\hline
\endfoot
& & & \\
\hline
\endlastfoot

$\mathfrak{t} + 2$ & $\mathfrak{t} + 2 - \frac{N}{2}$ &
$\left( \mathfrak{t} , 0 \right)^{-\mathfrak{t}}
\oplus \left( \mathfrak{t} - \frac{1}{2} , \frac{1}{2} \right)^{-\mathfrak{t}+1}$
 & $(2\mathfrak{t},0,0)_{SU(4)}$ \\[8pt]

 &  &
$\oplus \dots
\oplus \left( 0 , \mathfrak{t} \right)^{\mathfrak{t}}$
 & $(\underbrace{0,\dots,0}_{(N-1)},1)_{USp(2N)}$ \\[8pt]

\hline
& & & \\

$\mathfrak{t} + \frac{5}{2}$ & $\mathfrak{t} + 3 - \frac{N}{2}$ &
$\left( \mathfrak{t} + \frac{1}{2} , 0 \right)^{-\mathfrak{t}-\frac{1}{2}}
\oplus \left( \mathfrak{t} , \frac{1}{2} \right)^{-\mathfrak{t}+\frac{1}{2}}$
 & $(2\mathfrak{t}+1,0,0)_{SU(4)}$ \\[8pt]

 &  &
$\oplus \dots
\oplus \left( 0 , \mathfrak{t} + \frac{1}{2} \right)^{\mathfrak{t}+\frac{1}{2}}$
 & $(\underbrace{0,\dots,0}_{(N-2)},1,0)_{USp(2N)}$ \\[8pt]

\hline
& & & \\

\vdots & \vdots &
\qquad \vdots
 & \vdots \\[8pt]

\hline
& & & \\

$\mathfrak{t} + 2 + \frac{N}{2}$ & $\mathfrak{t} + 2 + \frac{N}{2}$ &
$\left( \mathfrak{t} + \frac{N}{2} , 0 \right)^{-\mathfrak{t}-\frac{N}{2}}$
 & $(2\mathfrak{t}+N,0,0)_{SU(4)}$ \\[8pt]

 &  &
$\oplus \left( \mathfrak{t} + \frac{N-1}{2} , \frac{1}{2} \right)^{-\mathfrak{t}-\frac{N-2}{2}}$
 & $(\underbrace{0,\dots,0}_N)_{USp(2N)}$ \\[8pt]

 &  &
$\oplus \dots
\oplus \left( 0 , \mathfrak{t} + \frac{N}{2} \right)^{\mathfrak{t}+\frac{N}{2}}$
 & \\[8pt]

\hline
\hline
& & & \\

$\mathfrak{t} + \frac{3}{2}$ & $\mathfrak{t} + 2 - \frac{N}{2}$ &
$\left( \mathfrak{t} - \frac{1}{2} , 0 \right)^{-\mathfrak{t}+\frac{1}{2}}$
 & $(2\mathfrak{t}-1,0,0)_{SU(4)}$ \\[8pt]

 &  &
$\oplus \left( \mathfrak{t} - 1 , \frac{1}{2} \right)^{-\mathfrak{t}+\frac{3}{2}}$
 & $(\underbrace{0,\dots,0}_{(N-2)},1,0)_{USp(2N)}$ \\[8pt]

 &  &
$\oplus \dots
\oplus \left( 0 , \mathfrak{t} - \frac{1}{2} \right)^{\mathfrak{t}-\frac{1}{2}}$
 & \\[8pt]

\hline
& & & \\

$\mathfrak{t} + 2$ & $\mathfrak{t} + 3 - \frac{N}{2}$ &
$\left( \mathfrak{t} , 0 \right)^{-\mathfrak{t}}
\oplus \left( \mathfrak{t} - \frac{1}{2} , \frac{1}{2} \right)^{-\mathfrak{t}+1}$
 & $(2\mathfrak{t},0,0)_{SU(4)}$ \\[8pt]

 &  &
$\oplus \dots
\oplus \left( 0 , \mathfrak{t} \right)^{\mathfrak{t}}$
 & $(\underbrace{0,\dots,0}_{(N-3)},1,0,0)_{USp(2N)}$ \\[8pt]

\hline
& & & \\

\vdots & \vdots &
\qquad \vdots
 & \vdots \\[8pt]

\hline
& & & \\

$\mathfrak{t} + 1 + \frac{N}{2}$ & $\mathfrak{t} + 1 + \frac{N}{2}$ &
$\left( \mathfrak{t} + \frac{N-2}{2} , 0 \right)^{-\mathfrak{t}-\frac{N-2}{2}}$
 & $(2\mathfrak{t}-2+N,0,0)_{SU(4)}$ \\[8pt]

 &  &
$\oplus \left( \mathfrak{t} + \frac{N-3}{2} , \frac{1}{2} \right)^{-\mathfrak{t}-\frac{N-4}{2}}$
 & $(\underbrace{0,\dots,0}_N)_{USp(2N)}$ \\[8pt]

 &  &
$\oplus \dots
\oplus \left( 0 , \mathfrak{t} + \frac{N-2}{2} \right)^{\mathfrak{t}+\frac{N-2}{2}}$
 & \\[8pt]

\hline
\hline
& & & \\

\vdots & \vdots &
\qquad \vdots
 & \vdots \\[8pt]

\vdots & \vdots &
\qquad \vdots
 & \vdots \\[8pt]

\hline
\hline
& & & \\

$\mathfrak{t} + 2 - \frac{n}{2}$ & $\mathfrak{t} + 2 - \frac{N}{2}$ &
$\left( \mathfrak{t} - \frac{n}{2} , 0 \right)^{-\mathfrak{t}+\frac{n}{2}}$
 & $(2\mathfrak{t}-n,0,0)_{SU(4)}$ \\[8pt]

 &  &
$\oplus \left( \mathfrak{t} - \frac{n+1}{2} , \frac{1}{2} \right)^{-\mathfrak{t}+\frac{n+2}{2}}$
 & $(\underbrace{0,\dots,0}_{(N-n-1)},1,0,\dots,0)_{USp(2N)}$ \\[8pt]

 &  &
$\oplus \dots
\oplus \left( 0 , \mathfrak{t} - \frac{n}{2} \right)^{\mathfrak{t} - \frac{n}{2}}$
 & \\[8pt]

\hline
& & & \\

$\mathfrak{t} + \frac{5}{2} - \frac{n}{2}$ & $\mathfrak{t} + 3 - \frac{N}{2}$ &
$\left( \mathfrak{t} - \frac{n-1}{2} , 0 \right)^{-\mathfrak{t}+\frac{n-1}{2}}$
 & $(2\mathfrak{t}-n+1,0,0)_{SU(4)}$ \\[8pt]

 &  &
$\oplus \left( \mathfrak{t} - \frac{n}{2} , \frac{1}{2} \right)^{-\mathfrak{t}+\frac{n+1}{2}}$
 & $(\underbrace{0,\dots,0}_{(N-n-2)},1,0,\dots,0)_{USp(2N)}$ \\[8pt]

 &  &
$\oplus \dots
\oplus \left( 0 , \mathfrak{t} - \frac{n-1}{2} \right)^{\mathfrak{t}-\frac{n-1}{2}}$
 & \\[8pt]

\hline
& & & \\

\vdots & \vdots &
\qquad \vdots
 & \vdots \\[8pt]

\hline
& & & \\

$\mathfrak{t} + 2 - n + \frac{N}{2}$ & $\mathfrak{t} + 2 - n + \frac{N}{2}$ &
$\left( \mathfrak{t} - n + \frac{N}{2} , 0 \right)^{-\mathfrak{t}+n-\frac{N}{2}}$
 & $(2\mathfrak{t}-2n+N,0,0)_{SU(4)}$ \\[8pt]

 &  &
$\oplus \left( \mathfrak{t} - n + \frac{N-1}{2} , \frac{1}{2} \right)^{-\mathfrak{t}+n-\frac{N-2}{2}}$
 & $(\underbrace{0,\dots,0}_N)_{USp(2N)}$ \\[8pt]

 &  &
$\oplus \dots
\oplus \left( 0 , \mathfrak{t} - n + \frac{N}{2} \right)^{\mathfrak{t}-n+\frac{N}{2}}$
 & \\[8pt]

\hline
\hline
& & & \\

\vdots & \vdots &
\qquad \vdots
 & \vdots \\[8pt]

\vdots & \vdots &
\qquad \vdots
 & \vdots \\[8pt]

\hline
\hline
& & & \\

$\mathfrak{t} + 2 - \frac{N}{2}$ & $\mathfrak{t} + 2 - \frac{N}{2}$ &
$\left( \mathfrak{t} - \frac{N}{2} , 0 \right)^{-\mathfrak{t}+\frac{N}{2}}$
 & $(2\mathfrak{t}-N,0,0)_{SU(4)}$ \\[8pt]

 &  &
$\oplus \left( \mathfrak{t} - \frac{N+1}{2} , \frac{1}{2} \right)^{-\mathfrak{t}+\frac{N+2}{2}}$
 & $(\underbrace{0,\dots,0}_N)_{USp(2N)}$ \\[8pt]

 &  &
$\oplus \dots
\oplus \left( 0 , \mathfrak{t} - \frac{N}{2} \right)^{\mathfrak{t}-\frac{N}{2}}$
 & \\[8pt]

\end{longtable}
\end{small}


\section{Deformed Minimal Unitary Supermultiplets of $OSp(8^*|4)$}
\label{deformedminrepsupermultipletsN=2}

Due to its importance as the symmetry superalgebra of the $S^4$
compactification of the eleven dimensional supergravity, we shall discus the
results for the case $N = 2$, i.e $OSp(8^*|4)$,  in more detail. The indices $r,s,\dots$ of $\alpha$-
and $\beta$-type (supersymmetry) fermionic oscillators  take the values 1,2 in this case.

Recall that the undeformed minimal unitary supermultiplet is obtained by taking the deformation parameter $\mathfrak{t} = 0$, which we present in Table \ref{Table:minrepsupermultipletN=2}.

\begin{small}
\begin{longtable}[c]{|c|c||l||c|c|}
\kill

\caption[The minimal unitary supermultiplet of $\mathfrak{osp}(8^*|4)$]
{The minimal unitary supermultiplet of $\mathfrak{osp}(8^*|4)$ defined by the lowest weight vector $\ket{\Omega^{(3/2)} \,,\, 1}$, which corresponds to the deformation parameter $\mathfrak{t} = 0$.
The decomposition of $SU(4)$ irreps with respect to $SU(2)_\mathcal{T} \times SU(2)_A
\times U(1)_\mathcal{J}$ is denoted by
$\left( \mathfrak{t} , \mathfrak{a} \right)^{\mathfrak{J}}$. $\mathcal{H}$ is the $AdS$
energy (negative conformal dimension), and $\mathfrak{H}$ is the total energy.
The Dynkin labels of the lowest energy $SU(4)$ representations of $SO^*(8)$ coincide with the Dynkin labels of the corresponding massless $6D$ conformal fields under the Lorentz group $SU^*(4)$. $USp(4)$ Dynkin labels of these fields are also given.
\label{Table:minrepsupermultipletN=2}} \\
\hline
& & & & \\
$\mathcal{H} = - \ell$ & $\mathfrak{H}$ &
$( \mathfrak{t} , \mathfrak{a} )^{\mathfrak{J}}$ & $SU(4)=SU^*(4)$ & $USp(4)$ \\
 &  &
 & Dynkin & Dynkin \\
& & & & \\
\hline
& & & & \\
\endfirsthead
\caption[]{(continued)} \\
\hline
& & & & \\
$\mathcal{H} = - \ell$ & $\mathfrak{H}$ &
$( \mathfrak{t} , \mathfrak{a} )^{\mathfrak{J}}$ & $SU(4)=SU^*(4)$ & $USp(4)$ \\
 &  &
 & Dynkin & Dynkin \\
& & & & \\
\hline
& & & & \\
\endhead
& & & & \\
\hline
\endfoot
& & & & \\
\hline
\endlastfoot

2 & 1 &
$(0,0)^0$
 & (0,0,0) & (0,1) \\[8pt]

\hline
& & & & \\

$\frac{5}{2}$ & 2 &
$\left( \frac{1}{2} , 0 \right)^{-\frac{1}{2}}
\oplus \left( 0 , \frac{1}{2} \right)^{+\frac{1}{2}}$
 & (1,0,0) & (1,0) \\[8pt]

\hline
& & & & \\

3 & 3 &
$\left( 1 , 0 \right)^{-1}
\oplus \left( \frac{1}{2} , \frac{1}{2} \right)^{0}
\oplus \left( 0 , 1 \right)^{+1}$
 & (2,0,0) & (0,0) \\[8pt]

\end{longtable}
\end{small}

The simplest deformed case is when the deformation parameter $\mathfrak{t} =g=
\frac{1}{2}$. This makes $\alpha_\mathfrak{t} = \frac{5}{2}$. In this case, we must
choose only one pair of deformation fermions  $\xi^1$ and $\chi^1$ (i.e. $P =
1$). Then we act on states
\begin{equation}
\chi^1 \ket{\psi_0^{(5/2)}} \oplus
\xi^1 \ket{\psi_0^{(5/2)}}
\end{equation}
with the coset generators $SU(4|N) \,/\, \left[ SU(2)_\mathcal{T} \times SU(2)_A \times
U(1)_\mathcal{J} \right]$ of grade zero subspace to
obtain a set of lowest energy states $\ket{\Omega^{(5/2)} \,,\, \sonebox}$ transforming in an irreducible
representation of $SU(4|2)$ with the Young supertableau ~$\sonebox$.
Repeatedly acting on these states with the supersymmetry generators in grade $+1$
subspace $\mathfrak{C}^+_D$, one obtains the  supermultiplet  given in Table \ref{Table:deformedsupermulN=2}.

This supermultiplet coincides with the doubleton supermultiplet given in \cite{Gunaydin:1999ci,Fernando:2001ak} with  the lowest energy K-type whose supertableu with respect to $SU(4|2)$ is $| \,\, \sonebox \, \rangle$.

\begin{small}
\begin{longtable}[c]{|c|c||l||c|c|}
\kill

\caption[The deformed minimal unitary supermultiplet of
$\mathfrak{osp}(8^*|4)$ corresponding to the deformation parameter $\mathfrak{t} =g= 1/2$]
{The deformed minimal unitary supermultiplet of $\mathfrak{osp}(8^*|4)$
corresponding to the deformation parameter $\mathfrak{t} = g = 1/2$. The decomposition of
$SU(4)$ irreps with respect to $SU(2)_\mathcal{T} \times SU(2)_A \times U(1)_\mathcal{J}$ is
denoted by $( \mathfrak{t} , \mathfrak{a} )^{\mathfrak{J}}$. $\mathcal{H}$ is the $AdS$
energy (negative conformal dimension), and $\mathfrak{H}$ is the total energy.
The $SU(4)$ Dynkin labels and the $USp(4)$ Dynkin labels are also given. The Dynkin labels of the lowest energy $SU(4)$ representations of $SO^*(8)$ coincide with the Dynkin labels of the corresponding massless $6D$ conformal fields under the Lorentz group $SU^*(4)$.
\label{Table:deformedsupermulN=2}} \\
\hline
& & & & \\
$\mathcal{H} = - \ell$ & $\mathfrak{H}$ &
$( \mathfrak{t} , \mathfrak{a} )^{\mathfrak{J}}$ & $SU(4)$ & $USp(4)$ \\
 &  &
 & Dynkin & Dynkin \\
& & & & \\
\hline
& & & & \\
\endfirsthead
\caption[]{(continued)} \\
\hline
& & & & \\
$\mathcal{H} = - \ell$ & $\mathfrak{H}$ &
$( \mathfrak{t} , \mathfrak{a} )^{\mathfrak{J}}$ & $SU(4)$ & $USp(4)$ \\
 &  &
 & Dynkin & Dynkin \\
& & & & \\
\hline
& & & & \\
\endhead
& & & & \\
\hline
\endfoot
& & & & \\
\hline
\endlastfoot

$\frac{5}{2}$ & $\frac{3}{2}$ &
$\left( \frac{1}{2} , 0 \right)^{-\frac{1}{2}}
\oplus \left( 0 , \frac{1}{2} \right)^{+\frac{1}{2}}$
 & (1,0,0) & (0,1) \\[8pt]

\hline
& & & & \\

3 & $\frac{5}{2}$ &
$\left( 1 , 0 \right)^{-1}
\oplus \left( \frac{1}{2} , \frac{1}{2} \right)^{0}
\left( 0 , 1 \right)^{+1}$
 & (2,0,0) & (1,0) \\[8pt]

\hline
& & & & \\

$\frac{7}{2}$ & $\frac{7}{2}$ &
$\left( \frac{3}{2} , 0 \right)^{-\frac{3}{2}}
\oplus \left( 1 , \frac{1}{2} \right)^{-\frac{1}{2}}
\oplus \left( \frac{1}{2} , 1 \right)^{+\frac{1}{2}}
\oplus \left( 0 , \frac{3}{2} \right)^{+\frac{3}{2}}$
 & (3,0,0) & (0,0) \\[8pt]

\hline
& & & & \\

2 & $\frac{3}{2}$ &
$\left( 0 , 0 \right)^0$
 & (0,0,0) & (1,0) \\[8pt]

\hline
& & & & \\

$\frac{5}{2}$ & $\frac{5}{2}$ &
$\left( \frac{1}{2} , 0 \right)^{-\frac{1}{2}}
\oplus \left( 0 , \frac{1}{2} \right)^{+\frac{1}{2}}$
 & (1,0,0) & (0,0) \\[8pt]

\end{longtable}
\end{small}

All higher spin  doubleton supermultiplets can be obtained similarly as deformations of the minimal unitary supermutiplet by
choosing the deformation parameter $\mathfrak{t}=g=P/2$ to take on all allowed values $\mathfrak{t} = 1/2,1,3/2,\dots$. The resulting    general deformed  supermultiplet (higher spin doubleton) is give in Table \ref{Table:gendeformedsupermulN=2}.

This supermultiplet matches exactly the corresponding doubleton
supermultiplet with lowest energy K-type
$| \,\, \underbrace{\sgenrowbox}_{2 \mathfrak{t}=2g=P} \,\, \rangle$ given  in
\cite{Gunaydin:1999ci,Fernando:2001ak}.

\begin{small}
\begin{longtable}[c]{|c|c||l||c|c|}
\kill

\caption[The general deformed minimal unitary supermultiplet of
$\mathfrak{osp}(8^*|4)$ corresponding to the deformation parameter $\mathfrak{t}=g=P/2$]
{The general deformed minimal unitary supermultiplet of
$\mathfrak{osp}(8^*|4)$ corresponding to the deformation parameter $\mathfrak{t} = g = P/2$.
The decomposition of $SU(4)$ irreps with respect to $SU(2)_\mathcal{T} \times SU(2)_A
\times U(1)_\mathcal{J}$ is denoted by $\left( \mathfrak{t} , \mathfrak{a}
\right)^{\mathfrak{J}}$. $\mathcal{H}$ is the $AdS$ energy (negative conformal
dimension), and $\mathfrak{H}$ is the total energy. The $SU(4)$ Dynkin labels
and the $USp(4)$ Dynkin labels are also given. The Dynkin labels of the lowest energy $SU(4)$ representations of $SO^*(8)$ coincide with the Dynkin labels of the corresponding massless $6D$ conformal fields under the Lorentz group $SU^*(4)$.
\label{Table:gendeformedsupermulN=2}} \\
\hline
& & & & \\
$\mathcal{H} = - \ell$ & $\mathfrak{H}$ &
$( \mathfrak{t} , \mathfrak{a} )^{\mathfrak{J}}$ & $SU(4)$ & $USp(4)$ \\
 &  &
 & Dynkin & Dynkin \\
& & & & \\
\hline
& & & & \\
\endfirsthead
\caption[]{(continued)} \\
\hline
& & & & \\
$\mathcal{H} = - \ell$ & $\mathfrak{H}$ &
$( \mathfrak{t} , \mathfrak{a} )^{\mathfrak{J}}$ & $SU(4)$ & $USp(4)$ \\
 &  &
 & Dynkin & Dynkin \\
& & & & \\
\hline
& & & & \\
\endhead
& & & & \\
\hline
\endfoot
& & & & \\
\hline
\endlastfoot

$\mathfrak{t} + 2$ & $\mathfrak{t} + 1$ &
$\left( \mathfrak{t} , 0 \right)^{-\mathfrak{t}}
\oplus \left( \mathfrak{t} - \frac{1}{2} , \frac{1}{2} \right)^{-\mathfrak{t}+1}
\oplus \dots \dots$
 & $(2\mathfrak{t},0,0)$ & (0,1) \\

 &  &
$\dots \dots
\oplus \left( 0 , \mathfrak{t} \right)^{+\mathfrak{t}}$
 &  &  \\[8pt]

\hline
& & & & \\

$\mathfrak{t} + \frac{5}{2}$ & $\mathfrak{t} + 2$ &
$\left( \mathfrak{t} + \frac{1}{2} , 0 \right)^{-\mathfrak{t}-\frac{1}{2}}
\oplus \left( \mathfrak{t} , \frac{1}{2} \right)^{-\mathfrak{t}+\frac{1}{2}}
\oplus \dots \dots$
 & $(2\mathfrak{t}+1,0,0)$ & (1,0) \\

 &  &
$\dots \dots
\oplus \left( 0 , \mathfrak{t} + \frac{1}{2} \right)^{\mathfrak{t}+\frac{1}{2}}$
 &  &  \\[8pt]

\hline
& & & & \\

$\mathfrak{t} + 3$ & $\mathfrak{t} + 3$ &
$\left( \mathfrak{t} + 1 , 0 \right)^{-\mathfrak{t}-1}
\oplus \left( \mathfrak{t} + \frac{1}{2} , \frac{1}{2} \right)^{-\mathfrak{t}-\frac{1}{2}}
\oplus \dots \dots$
 & $(2\mathfrak{t}+2,0,0)$ & (0,0) \\

 &  &
$\dots \dots
\oplus \left( 0 , \mathfrak{t} + 1 \right)^{\mathfrak{t}+1}$
 &  &  \\[8pt]

\hline
& & & & \\

$\mathfrak{t} + \frac{3}{2}$ & $\mathfrak{t} + 1$ &
$\left( \mathfrak{t} - \frac{1}{2} , 0 \right)^{-\mathfrak{t}+\frac{1}{2}}
\oplus \left( \mathfrak{t} - 1 , \frac{1}{2} \right)^{-\mathfrak{t}+\frac{3}{2}}
\oplus \dots \dots$
 & $(2\mathfrak{t}-1,0,0)$ & (1,0) \\

 &  &
$\dots \dots
\oplus \left( 0 , \mathfrak{t} - \frac{1}{2} \right)^{\mathfrak{t}-\frac{1}{2}}$
 &  &  \\[8pt]

\hline
& & & & \\

$\mathfrak{t} + 2$ & $\mathfrak{t} + 2$ &
$\left( \mathfrak{t} , 0 \right)^{-\mathfrak{t}}
\oplus \left( \mathfrak{t} - \frac{1}{2} , \frac{1}{2} \right)^{-\mathfrak{t}+1}
\oplus \dots \dots$
 & $(2\mathfrak{t},0,0)$ & (0,0) \\

 &  &
$\dots \dots
\oplus \left( 0 , \mathfrak{t} \right)^{+\mathfrak{t}}$
 &  &  \\[8pt]

\hline
& & & & \\

$\mathfrak{t} + 1$ & $\mathfrak{t} + 1$ &
$\left( \mathfrak{t} - 1 , 0 \right)^{-\mathfrak{t}+1}
\oplus \left( \mathfrak{t} - \frac{1}{2} , \frac{1}{2} \right)^{-\mathfrak{t}+2}
\oplus \dots \dots$
 & $(2\mathfrak{t}-2,0,0)$ & (0,0) \\

 &  &
$\dots \dots
\oplus \left( 0 , \mathfrak{t} - 1 \right)^{\mathfrak{t}-1}$
 &  &  \\[8pt]

\end{longtable}
\end{small}


{\bf Acknowledgements:} We would like to thank Oleksandr Pavlyk for many stimulating discussions and his generous help with Mathematica. S.F. would like to thank the Center for Fundamental Theory of  the
Institute for Gravitation and the Cosmos at Pennsylvania State University, where part of this work was done, for their warm hospitality. \\
  This work was supported in part by the National
Science Foundation under grants numbered PHY-0555605 and PHY-0855356. Any opinions,
findings and conclusions or recommendations expressed in this
material are those of the authors and do not necessarily reflect the
views of the National Science Foundation.


\appendix

\numberwithin{equation}{section}

\section*{Appendix}


\section{Construction of Finite-Dimensional Representations of $USp(2N)$ in
terms of Fermionic Oscillators}
\label{USp(2N)}

To realize the generators of the compact Lie algebra $\mathfrak{usp}(2N)$, we
define two new sets of $N$ fermionic oscillators $\alpha_r$, $\beta_r$ and
their hermitian conjugates $\alpha^r = \left( \alpha_r \right)^\dag$,
$\beta^r = \left( \beta_r \right)^\dag$ ($r = 1,2,\dots,N$), such that they
satisfy the usual anti-commutation relations:
\begin{equation}
\anticommute{\alpha_r}{\alpha^s}
= \anticommute{\beta_r}{\beta^s}
= \delta^s_r
\qquad \qquad
\anticommute{\alpha_r}{\alpha_s}
= \anticommute{\alpha_r}{\beta_s}
= \anticommute{\beta_r}{\beta_s}
= 0
\label{susyfermions}
\end{equation}

The Lie algebra $\mathfrak{usp}(2N)$ has a 3-graded decomposition with
respect to its subalgebra $\mathfrak{u}(N)$ as follows:
\begin{equation}
\begin{split}
\mathfrak{usp}(2N)
&= \mathfrak{g}^{(-1)} \, \oplus \, \mathfrak{g}^{(0)} \, \oplus \,
\mathfrak{g}^{(+1)}
\\
&= \, S_{rs} \, \oplus \, M^r_{~s} \, \oplus \, S^{rs}
\end{split}
\end{equation}
where the generators $M^r_{~s}$ form the $\mathfrak{u}(N)$ subalgebra. They can be realized as bilinears of the fermionic oscillators:\footnote{Note that realizing the generators of $USp(2N)$ as bilinears of a single pair of fermionic oscillators leads to a finite  set of irreps, which are the compact analogs of "doubleton" irreps. To construct more general irreps of $USp(2N)$ one needs to take an arbitrary number (color) of pairs of these oscillators and sum over the color index. See \cite{Gunaydin:1990ag} for a general treatment.}
\begin{equation}
\begin{split}
S_{rs}
&= \alpha_r \beta_s + \alpha_s \beta_r
\\
M^r_{~s}
&= \alpha^r \alpha_s - \beta_s \beta^r
\\
S^{rs}
&= \beta^r \alpha^s + \beta^s \alpha^r
 = \left( S_{rs} \right)^\dag \,.
\end{split}
\label{USp(2N)_gen}
\end{equation}
 The
$\mathfrak{usp}(2N)$ generators satisfy the following commutation relations:
\begin{equation}
\begin{split}
\commute{S_{rs}}{S^{tu}}
&= - \delta^t_s \, M^u_{~r} - \delta^t_r \, M^u_{~s}
   - \delta^u_s \, M^t_{~r} - \delta^u_r \, M^t_{~s}
\\
\commute{M^r_{~s}}{S_{tu}}
&= - \delta^r_u \, S_{st} - \delta^r_t \, S_{su}
\\
\commute{M^r_{~s}}{S^{tu}}
&= \delta^u_s \, S^{rt} + \delta^t_s \, S^{ru}
\\
\commute{M^r_{~s}}{M^t_{~u}}
&= \delta^t_s \, M^r_{~u} - \delta^r_u \, M^t_{~s}
\end{split}
\end{equation}
The quadratic Casimir of $\mathfrak{usp}(2N)$ is given by
\begin{equation}
\begin{split}
\mathcal{C}_2 \left[ \mathfrak{usp}(2N) \right]
&= M^r_{~s} M^s_{~r}
   + \frac{1}{2} \left( S_{rs} S^{rs} + S^{rs} S_{rs} \right)
\\
&= N \left( N + 2 \right)
   - \left( N_\alpha + N_\beta \right)
     \left[ \left( N_\alpha + N_\beta \right) + 2 \right]
   - 8 \, \alpha^{(r} \beta^{s)} \, \alpha_{(r} \beta_{s)}
\end{split}
\end{equation}
where ``$(rs)$'' represents symmetrization of weight one, $\alpha_{(r}
\beta_{s)} = \frac{1}{2} \left( \alpha_r \beta_s + \alpha_s \beta_r \right)$.

We choose the Fock vacuum of these fermionic oscillators such that
\begin{equation}
\alpha_r \ket{0}_F
= \beta_r \ket{0}_F
= 0 \,.
\end{equation}

To generate an irrep of $USp(2N)$ in this Fock space in a $U(N)$ basis, one
chooses a set of states $\ket{\Omega}$, transforming irreducibly under $U(N)$
and is annihilated by all grade $-1$ generators $S_{rs}$, and act on it with
grade $+1$ generators $S^{rs}$ \cite{Gunaydin:1990ag}.

The possible sets of states $\ket{\Omega}$, that transform irreducibly under
$U(N)$ and are annihilated by $S_{rs}$, are of the form
\begin{equation}
\alpha^{r_1} \alpha^{r_2} \dots \alpha^{r_m} \ket{0}_F
\end{equation}
or of the equivalent form
\begin{equation}
\beta^{r_1} \beta^{r_2} \dots \beta^{r_m} \ket{0}_F \,.
\end{equation}
where $m \le N$. They lead to irreps of $USp(2N)$ with Dynkin labels
\cite{Gunaydin:1990ag}
\begin{equation}
( \, \underbrace{0 , \dots , 0}_{(N-m-1)} , 1 ,
\underbrace{0 , \dots , 0}_{(m)} \, ) \,.
\end{equation}
In addition, we have the following states
\begin{equation}
\alpha^{[r} \beta^{s]} \ket{0}_F
= \frac{1}{2} \left( \alpha^r \beta^s - \alpha^s \beta^r \right) \ket{0}_F
\end{equation}
that are annihilated by all grade $-1$ generators $S_{tu}$. They lead to the
irrep of $USp(2N)$ with Dynkin labels
\begin{equation}
( \, \underbrace{0 , \dots , 0}_{(N-3)} , 1 , 0 , 0 \, ) \,.
\end{equation}
Note that in the special case of $\mathfrak{usp}(4)$, the states $\alpha^r
\alpha^s \ket{0}_F$, $\beta^r \beta^s \ket{0}_F$ and $\alpha^{[r} \beta^{s]}
\ket{0}_F$ all lead to the trivial representation.

Also note that the following bilinears of these $\alpha$- and $\beta$-type
fermionic oscillators:
\begin{equation}
F_+ = \alpha^r \beta_r
\qquad \qquad
F_- = \beta^r \alpha_r
\qquad \qquad
F_0 = \frac{1}{2} \left( N_\alpha - N_\beta \right)
\label{SU(2)F_gen}
\end{equation}
where $N_\alpha = \alpha^r \alpha_r$ and $N_\beta = \beta^r \beta_r$ are the
respective number operators, generate a $\mathfrak{usp}(2)_F \simeq
\mathfrak{su}(2)_F$ algebra
\begin{equation}
\commute{F_+}{F_-} = 2 \, F_0
\qquad \qquad \qquad
\commute{F_0}{F_\pm} = \pm F_\pm
\end{equation}
that commutes with the $\mathfrak{usp}(2N)$ algebra defined above.
Nonetheless, the equivalent irreps of $USp(2N)$ constructed from the states
$\ket{\Omega}$ involving only $\alpha$-type excitations or $\beta$-type
excitations can form non-trivial representations of this $USp(2)_F$.

For example, the two irreps labeled by $(1,0)$ of $USp(4)$ constructed from
$\alpha^r \ket{0}$ and $\beta^r \ket{0}$ form spin $\frac{1}{2}$
representation (doublet) of $USp(2)_F$. The three singlet irreps of
$USp(4)$ corrsponding to $\alpha^r \alpha^s \ket{0}$, $\beta^r \beta^s
\ket{0}$ and $\alpha^{[r} \beta^{s]} \ket{0}$ form the spin 1 representation
(triplet) of $USp(2)_F$. The irrep of $USp(4)$ with Dynkin labels $(0,1)$
defined by the vacuum state $\ket{\Omega} = \ket{0}$ is a singlet of
$USp(2)_F$.

We should stress that the representations of $USp(2N)$ obtained above by
using two sets of fermionic oscillators transforming in the fundamental
representation of the subgroup $U(N)$ are the compact analogs of the
doubleton representations of $SO^*(2M)$ constructed using two sets of bosonic
oscillators transforming in the fundamental representation of $U(M)$
\cite{Gunaydin:1990ag}. By realizing the generators of $USp(2N)$ in terms of
an arbitrary (even) number of sets of oscillators, one can construct all the
finite dimensional representations of $USp(2N)$
\cite{Gunaydin:1990ag,Gunaydin:1984wc}.


\providecommand{\href}[2]{#2}\begingroup\raggedright\endgroup


\begin{thebibliography}{10}

\bibitem{Gunaydin:1981dc}
M.~G{\"u}naydin and C.~Saclioglu, ``{Bosonic construction of the Lie algebras
  of some noncompact groups appearing in supergravity theories and their
  oscillator-like unitary representations},'' {\em Phys. Lett.} {\bf B108}
  (1982)
180.

\bibitem{Gunaydin:1981yq}
M.~G{\"u}naydin and C.~Saclioglu, ``{Oscillator-like unitary representations of
  noncompact groups with a Jordan structure and the noncompact groups of
  supergravity},'' {\em Commun. Math. Phys.} {\bf 87} (1982)
159.

\bibitem{Gunaydin:1981zm}
M.~G{\"u}naydin, ``{Unitary realizations of the noncompact symmetry groups of
  supergravity},''. Presented at 2nd Europhysics Study Conf. on Unification of
  Fundamental Interactions, Erice, Sicily, Oct 6-14, 1981.

\bibitem{Bars:1982ep}
I.~Bars and M.~Gunaydin, ``{Unitary Representations of Noncompact
  Supergroups},'' {\em Commun. Math. Phys.} {\bf 91} (1983)
31.

\bibitem{Gunaydin:1984fk}
M.~Gunaydin and N.~Marcus, ``{The Spectrum of the $S^5$ Compactification of the
  Chiral $N=2, D=10$ Supergravity and the Unitary Supermultiplets of $U(2,
  2/4)$},'' {\em Class. Quant. Grav.} {\bf 2} (1985)
L11.

\bibitem{Gunaydin:1985tc}
M.~Gunaydin and N.~P. Warner, ``{Unitary Supermultiplets of $OSp(8/4,R)$ and
  the Spectrum of the $S^7$ Compactification of Eleven-Dimensional
  Supergravity},'' {\em Nucl. Phys.} {\bf B272} (1986)
99.

\bibitem{Gunaydin:1984wc}
M.~Gunaydin, P.~van Nieuwenhuizen, and N.~P. Warner, ``{General Construction of
  the Unitary Representations of Anti-De Sitter Superalgebras and the Spectrum
  of the $S^4$ Compactification of Eleven-Dimensional Supergravity},'' {\em
  Nucl. Phys.} {\bf B255} (1985)
63.

\bibitem{Maldacena:1997re}
J.~M. Maldacena, ``{The large N limit of superconformal field theories and
  supergravity},'' {\em Adv. Theor. Math. Phys.} {\bf 2} (1998) 231--252,
\href{http://www.arXiv.org/abs/hep-th/9711200}{{\tt hep-th/9711200}}.

\bibitem{Witten:1998qj}
E.~Witten, ``{Anti-de Sitter space and holography},'' {\em Adv. Theor. Math.
  Phys.} {\bf 2} (1998) 253--291,
\href{http://www.arXiv.org/abs/hep-th/9802150}{{\tt hep-th/9802150}}.

\bibitem{Gubser:1998bc}
S.~S. Gubser, I.~R. Klebanov, and A.~M. Polyakov, ``{Gauge theory correlators
  from non-critical string theory},'' {\em Phys. Lett.} {\bf B428} (1998)
  105--114,
\href{http://www.arXiv.org/abs/hep-th/9802109}{{\tt hep-th/9802109}}.

\bibitem{MR0342049}
A.~Joseph, ``Minimal realizations and spectrum generating algebras,'' {\em
  Comm. Math. Phys.} {\bf 36} (1974) 325--338.

\bibitem{MR644845}
D.~A. Vogan, Jr., ``Singular unitary representations,'' in {\em Noncommutative
  harmonic analysis and {L}ie groups ({M}arseille, 1980)}, vol.~880 of {\em
  Lecture Notes in Math.}, pp.~506--535.
\newblock Springer, Berlin, 1981.

\bibitem{MR1103588}
B.~Kostant, ``The vanishing of scalar curvature and the minimal representation
  of {${SO}(4,4)$},'' in {\em Operator algebras, unitary representations,
  enveloping algebras, and invariant theory ({P}aris, 1989)}, vol.~92 of {\em
  Progr. Math.}, pp.~85--124.
\newblock Birkh\"auser Boston, Boston, MA, 1990.

\bibitem{MR1159103}
D.~Kazhdan and G.~Savin, ``The smallest representation of simply laced
  groups,'' in {\em Festschrift in honor of I. I. Piatetski-Shapiro on the
  occasion of his sixtieth birthday, Part I (Ramat Aviv, 1989)}, vol.~2 of {\em
  Israel Math. Conf. Proc.}, pp.~209--223.
\newblock Weizmann, Jerusalem, 1990.

\bibitem{MR1372999}
R.~Brylinski and B.~Kostant, ``Lagrangian models of minimal representations of
  {$E\sb 6$}, {$E\sb 7$} and {$E\sb 8$},'' in {\em Functional analysis on the
  eve of the 21st century, Vol.\ 1 (New Brunswick, NJ, 1993)}, vol.~131 of {\em
  Progr. Math.}, pp.~13--63.
\newblock Birkh\"auser Boston, Boston, MA, 1995.

\bibitem{MR1278630}
R.~Brylinski and B.~Kostant, ``Minimal representations, geometric quantization,
  and unitarity,'' {\em Proc. Nat. Acad. Sci. U.S.A.} {\bf 91} (1994), no.~13,
  6026--6029.

\bibitem{Kazhdan:2001nx}
D.~Kazhdan, B.~Pioline, and A.~Waldron, ``{M}inimal representations, spherical
  vectors, and exceptional theta series. {I},'' {\em Commun. Math. Phys.} {\bf
  226} (2002) 1--40,
\href{http://www.arXiv.org/abs/hep-th/0107222}{{\tt hep-th/0107222}}.

\bibitem{MR1327538}
B.~H. Gross and N.~R. Wallach, ``A distinguished family of unitary
  representations for the exceptional groups of real rank {$=4$},'' in {\em Lie
  theory and geometry}, vol.~123 of {\em Progr. Math.}, pp.~289--304.
\newblock Birkh\"auser Boston, Boston, MA, 1994.

\bibitem{MR1108044}
B.~Binegar and R.~Zierau, ``Unitarization of a singular representation of
  {${\rm SO}(p,q)$},'' {\em Comm. Math. Phys.} {\bf 138} (1991), no.~2,
  245--258.

\bibitem{MR2020550}
T.~Kobayashi and B.~{\O}rsted, ``Analysis on the minimal representation of
  {$O(p,q)$}. {I}. {R}ealization via conformal geometry,'' {\em Adv. Math.}
  {\bf 180} (2003), no.~2, 486--512.

\bibitem{MR2020551}
T.~Kobayashi and B.~{\O}rsted, ``Analysis on the minimal representation of {$
  O(p,q)$}. {II}. {B}ranching laws,'' {\em Adv. Math.} {\bf 180} (2003), no.~2,
  513--550.

\bibitem{MR2020552}
T.~Kobayashi and B.~{\O}rsted, ``Analysis on the minimal representation of {$
  O(p,q)$}. {III}. {U}ltrahyperbolic equations on {${\mathbf{
  R}}^{p-1,q-1}$},'' {\em Adv. Math.} {\bf 180} (2003), no.~2, 551--595.

\bibitem{Gover:2009vc}
A.~R. Gover and A.~Waldron, ``{The so(d+2,2) Minimal Representation and Ambient
  Tractors: the Conformal Geometry of Momentum Space},''
\href{http://www.arXiv.org/abs/0903.1394}{{\tt 0903.1394}}.

\bibitem{Ferrara:1997uz}
S.~Ferrara and M.~G\"unaydin, ``{O}rbits of exceptional groups, duality and
  {BPS} states in string theory,'' {\em Int. J. Mod. Phys.} {\bf A13} (1998)
  2075--2088,
\href{http://www.arXiv.org/abs/hep-th/9708025}{{\tt hep-th/9708025}}.

\bibitem{Gunaydin:2000xr}
M.~G{\"u}naydin, K.~Koepsell, and H.~Nicolai, ``{C}onformal and quasiconformal
  realizations of exceptional {L}ie groups,'' {\em Commun. Math. Phys.} {\bf
  221} (2001) 57--76,
\href{http://www.arXiv.org/abs/hep-th/0008063}{{\tt hep-th/0008063}}.

\bibitem{Gunaydin:2004ku}
M.~G{\"u}naydin, ``{Realizations of exceptional U-duality groups as conformal
  and quasiconformal groups and their minimal unitary representations},'' {\em
  Comment. Phys. Math. Soc. Sci. Fenn.} {\bf 166} (2004) 111--125,
\href{http://www.arXiv.org/abs/hep-th/0409263}{{\tt hep-th/0409263}}.

\bibitem{Gunaydin:2003qm}
M.~G{\"u}naydin, ``{Realizations of exceptional U-duality groups as conformal
  and quasi-conformal groups and their minimal unitary representations},''.
  Prepared for 3rd International Symposium on Quantum Theory and Symmetries
  (QTS3), Cincinnati, Ohio, 10-14 Sep 2003.

\bibitem{Gunaydin:2005gd}
M.~G{\"u}naydin, ``{U}nitary realizations of {U}-duality groups as conformal
  and quasiconformal groups and extremal black holes of supergravity
  theories,'' {\em AIP Conf. Proc.} {\bf 767} (2005) 268--287,
\href{http://www.arXiv.org/abs/hep-th/0502235}{{\tt hep-th/0502235}}.

\bibitem{Gunaydin:2009pk}
M.~Gunaydin, ``{Lectures on Spectrum Generating Symmetries and U-duality in
  Supergravity, Extremal Black Holes, Quantum Attractors and Harmonic
  Superspace},''
\href{http://www.arXiv.org/abs/0908.0374}{{\tt 0908.0374}}.

\bibitem{Gunaydin:2005mx}
M.~G{\"u}naydin, A.~Neitzke, B.~Pioline, and A.~Waldron, ``{BPS} black holes,
  quantum attractor flows and automorphic forms,'' {\em Phys. Rev.} {\bf D73}
  (2006) 084019,
\href{http://www.arXiv.org/abs/hep-th/0512296}{{\tt hep-th/0512296}}.

\bibitem{Gunaydin:2007bg}
M.~G{\"u}naydin, A.~Neitzke, B.~Pioline, and A.~Waldron, ``{Quantum Attractor
  Flows},'' {\em JHEP} {\bf 09} (2007) 056,
\href{http://www.arXiv.org/abs/0707.0267}{{\tt 0707.0267}}.

\bibitem{Gunaydin:2007qq}
M.~G{\"u}naydin, A.~Neitzke, O.~Pavlyk, and B.~Pioline, ``{Quasi-conformal
  actions, quaternionic discrete series and twistors: ${SU(2,1)}$ and
  ${G_{2(2)}}$},'' {\em Commun. Math. Phys.} {\bf 283} (2008) 169--226,
\href{http://www.arXiv.org/abs/0707.1669}{{\tt 0707.1669}}.

\bibitem{Breitenlohner:1987dg}
P.~Breitenlohner, G.~W. Gibbons, and D.~Maison, ``{F}our-dimensional black
  holes from {K}aluza-{K}lein theories,'' {\em Commun. Math. Phys.} {\bf 120}
  (1988)
295.

\bibitem{Gunaydin:2005zz}
M.~G{\"u}naydin and O.~Pavlyk, ``Generalized spacetimes defined by cubic forms
  and the minimal unitary realizations of their quasiconformal groups,'' {\em
  JHEP} {\bf 08} (2005) 101,
\href{http://www.arXiv.org/abs/hep-th/0506010}{{\tt hep-th/0506010}}.

\bibitem{Gunaydin:2001bt}
M.~G{\"u}naydin, K.~Koepsell, and H.~Nicolai, ``{T}he minimal unitary
  representation of ${E_{8(8)}}$,'' {\em Adv. Theor. Math. Phys.} {\bf 5}
  (2002) 923--946,
\href{http://www.arXiv.org/abs/hep-th/0109005}{{\tt hep-th/0109005}}.

\bibitem{Gunaydin:1983rk}
M.~G{\"u}naydin, G.~Sierra, and P.~K. Townsend, ``Exceptional supergravity
  theories and the magic square,'' {\em Phys. Lett.} {\bf B133} (1983)
72.

\bibitem{Gunaydin:2004md}
M.~G{\"u}naydin and O.~Pavlyk, ``{M}inimal unitary realizations of exceptional
  {U}-duality groups and their subgroups as quasiconformal groups,'' {\em JHEP}
  {\bf 01} (2005) 019,
\href{http://www.arXiv.org/abs/hep-th/0409272}{{\tt hep-th/0409272}}.

\bibitem{Gunaydin:2006vz}
M.~G{\"u}naydin and O.~Pavlyk, ``A unified approach to the minimal unitary
  realizations of noncompact groups and supergroups,'' {\em JHEP} {\bf 09}
  (2006) 050,
\href{http://www.arXiv.org/abs/hep-th/0604077}{{\tt hep-th/0604077}}.

\bibitem{Gunaydin:1988kz}
M.~Gunaydin and S.~J. Hyun, ``{Unitary lowest weight representations of the
  noncompact supergroup $OSp(2n|2m,R)$},'' {\em J. Math. Phys.} {\bf 29} (1988)
2367.

\bibitem{Gunaydin:1987hb}
M.~Gunaydin, ``{Unitary highest weight representations of noncompact
  supergroups},'' {\em J. Math. Phys.} {\bf 29} (1988)
1275--1282.

\bibitem{Fernando:2009fq}
S.~Fernando and M.~Gunaydin, ``{Minimal unitary representation of SU(2,2) and
  its deformations as massless conformal fields and their supersymmetric
  extensions},''
\href{http://www.arXiv.org/abs/0908.3624}{{\tt 0908.3624}}.

\bibitem{Gunaydin:1998sw}
M.~Gunaydin, D.~Minic, and M.~Zagermann, ``{4D doubleton conformal theories,
  CPT and II B string on AdS(5) x S(5)},'' {\em Nucl. Phys.} {\bf B534} (1998)
  96--120,
\href{http://www.arXiv.org/abs/hep-th/9806042}{{\tt hep-th/9806042}}.

\bibitem{Gunaydin:1998jc}
M.~Gunaydin, D.~Minic, and M.~Zagermann, ``{Novel supermultiplets of
  $SU(2,2|4)$ and the $AdS_5/CFT_4 $ duality},'' {\em Nucl. Phys.} {\bf B544}
  (1999) 737--758,
\href{http://www.arXiv.org/abs/hep-th/9810226}{{\tt hep-th/9810226}}.

\bibitem{Mack:1969dg}
G.~Mack and I.~Todorov, ``{Irreducibility of the ladder representations of
  U(2,2) when restricted to the Poincare subgroup},'' {\em J. Math. Phys.} {\bf
  10} (1969)
2078--2085.

\bibitem{Fernando:2010dp}
S.~Fernando and M.~Gunaydin, ``{Minimal unitary representation of SO*(8) =
  SO(6,2) and its SU(2) deformations as massless 6D conformal fields and their
  supersymmetric extensions},''
\href{http://www.arXiv.org/abs/1005.3580}{{\tt 1005.3580}}.

\bibitem{Gunaydin:1999ci}
M.~Gunaydin and S.~Takemae, ``{Unitary supermultiplets of $OSp(8*|4)$ and the
  AdS(7)/CFT(6) duality},'' {\em Nucl. Phys.} {\bf B578} (2000) 405--448,
\href{http://www.arXiv.org/abs/hep-th/9910110}{{\tt hep-th/9910110}}.

\bibitem{Fernando:2001ak}
S.~Fernando, M.~Gunaydin, and S.~Takemae, ``{Supercoherent states of
  $OSp(8^*|2N)$, conformal superfields and the $AdS_7/CFT_6$ duality},'' {\em
  Nucl. Phys.} {\bf B628} (2002) 79--111,
\href{http://www.arXiv.org/abs/hep-th/0106161}{{\tt hep-th/0106161}}.

\bibitem{Ferrara:2000xg}
S.~Ferrara and E.~Sokatchev, ``{Representations of (1,0) and (2,0)
  superconformal algebras in six dimensions: Massless and short superfields},''
  {\em Lett. Math. Phys.} {\bf 51} (2000) 55--69,
\href{http://www.arXiv.org/abs/hep-th/0001178}{{\tt hep-th/0001178}}.

\bibitem{Ferrara:2000dv}
S.~Ferrara and E.~Sokatchev, ``{Representations of superconformal algebras in
  the AdS(7/4)/CFT(6/3) correspondence},'' {\em J. Math. Phys.} {\bf 42} (2001)
  3015--3026,
\href{http://www.arXiv.org/abs/hep-th/0010117}{{\tt hep-th/0010117}}.

\bibitem{Minwalla:1997ka}
S.~Minwalla, ``{Restrictions imposed by superconformal invariance on quantum
  field theories},'' {\em Adv. Theor. Math. Phys.} {\bf 2} (1998) 781--846,
\href{http://www.arXiv.org/abs/hep-th/9712074}{{\tt hep-th/9712074}}.

\bibitem{Dobrev:2002dt}
V.~K. Dobrev, ``{Positive energy unitary irreducible representations of D = 6
  conformal supersymmetry},'' {\em J. Phys.} {\bf A35} (2002) 7079--7100,
\href{http://www.arXiv.org/abs/hep-th/0201076}{{\tt hep-th/0201076}}.

\bibitem{Gunaydin:1990ag}
M.~Gunaydin and R.~J. Scalise, ``{Unitary Lowest Weight Representations of the
  Noncompact Supergroup $OSp(2m*|2n)$},'' {\em J. Math. Phys.} {\bf 32} (1991)
599--606.

\bibitem{deAlfaro:1976je}
V.~de~Alfaro, S.~Fubini, and G.~Furlan, ``Conformal invariance in quantum
  mechanics,'' {\em Nuovo Cim.} {\bf A34} (1976)
569.

\bibitem{Casahorran:1995vt}
J.~Casahorran, ``{On a novel supersymmetric connection between harmonic and
  isotonic oscillators},'' {\em Physica A} {\bf 217} (1995) 429--39.
  DFTUZ-94-28.

\bibitem{carinena-2007}
J.~F. Carinena, A.~M. Perelomov, M.~F. Ranada, and M.~Santander, ``A quantum
  exactly solvable non-linear oscillator related with the isotonic
  oscillator,'' {\em J. Phys. A: Math. Theor.} {\bf 41} (2008) 085301.

\bibitem{MR858831}
A.~Perelomov, {\em Generalized coherent states and their applications}.
\newblock Texts and Monographs in Physics. Springer-Verlag, Berlin, 1986.

\end{thebibliography}
\end{document}